\documentclass[12pt]{iopart}

\usepackage{iopams}
\usepackage{graphicx}
\usepackage{amssymb}
\usepackage{dsfont}
\usepackage{bm}
\usepackage{epstopdf}

\usepackage{color}

\newcommand{\be}{\begin{equation}}
\newcommand{\ee}{\end{equation}}

\begin{document}

\title{Percolation assisted excitation transport in discrete-time quantum walks}

\author{M. \v Stefa\v n\'ak, J. Novotn\'y and I. Jex}

\address{Department of Physics, Faculty of Nuclear Sciences and Physical Engineering, Czech Technical University in Prague, B\v
rehov\'a 7, 115 19 Praha 1 - Star\'e M\v{e}sto, Czech Republic}

\date{\today}

\begin{abstract}
Coherent transport of excitations along chains of coupled quantum
systems represents an interesting problem with a number of
applications ranging from quantum optics to solar cell technology. A convenient tool for studying such processes are quantum walks. They allow to determine in a quantitative way all the process features. We study the survival probability and the transport efficiency on a simple, highly symmetric graph represented by a ring. The propagation of excitation is modeled by a discrete-time (coined) quantum walk. For a two-state
quantum walk, where the excitation (walker) has to leave its actual
position to the neighboring sites, the survival probability decays exponentially and the transport efficiency is unity. The decay rate of the survival probability can be estimated using the leading eigenvalue of the evolution operator. However, if the excitation is allowed to stay at its present position, i.e. the propagation is modeled
by a lazy quantum walk, then part of the wave-packet can be
trapped in the vicinity of the origin and never reaches the sink. In
such a case, the survival probability does not vanish and the excitation transport is not efficient. The dependency of the transport efficiency on the initial state is determined. Nevertheless, we show that for some lazy quantum walks dynamical percolations of the ring eliminate the trapping effect and
efficient excitation transport can be achieved.
\end{abstract}

\maketitle

\section{Introduction}

Quantum walks \cite{adz,meyer,fg} emerged as an extension of the
concept of a classical random walk to the evolution of a quantum
particle on a graph or lattice. They were applied to various
problems of quantum information, such as quantum search of an
unsorted database
\cite{skw,childs:search:2004,reitzner:search,vasek:search,childs:search:2014},
perfect state transfer
\cite{kendon:qw:pst,wojcik:qw:pst,barr:qw:pst,zhan:qw:pst,gedik:qw:pst},
graph isomorphism testing \cite{gamble,berry,rudiger} or finding
structural anomalies in graphs \cite{reitzner,hillery,cottrell}.
Moreover, quantum walks were shown to be universal tools for quantum
computation \cite{childs,Lovett}. Up to date quantum walks were
realized in a number of experiments utilizing optically trapped
atoms \cite{karski}, cold ions \cite{schmitz,zahringer} and photons
\cite{and:1d,broome,peruzzo:corelated:photons,two:photon:waveguide,sansoni,and:2dwalk:science,delayed:choice}.
Recently, an experimental realization of quantum walk on dynamically
percolated line was performed \cite{elster:perc}.

Quantum walks have been also intensively studied in the context of
coherent transport on graphs or networks of coupled quantum systems.
Among the first studies the Hadamard walk on a half-line with one
absorbing boundary (sink) was analyzed in \cite{ambainis:absorp}. The authors have found that
the particle has a non-zero chance to escape to infinity, in
contrast to the classical result where the particle falls into the
sink with certainty. This problem was later studied in a number of
alternated configurations, coins a geometries
\cite{konno:absorp,bach:absorp,yamasaki:absorp,kwek:absorp,chandra:absorp,janos:absorp}. Dependence of the absorption probability on the
initial condition for quantum walk on finite line with two sinks was
analyzed in \cite{konno:absorp}. The time dependence of the
absorption probability was studied in \cite{bach:absorp} for quantum
walk on a line with one and two sinks. The authors have also proposed an
extension to $d$-dimensional walk with $d-1$-dimensional absorbing
wall. The paper \cite{yamasaki:absorp} analyzed the absorbing times
for quantum walks on hypercube. In \cite{kwek:absorp} the authors
have considered  quantum walk on a line with one or two moving
sinks. More recently, enhancement of transport by noise in quantum
walk on a closed loop was identified in \cite{chandra:absorp}. In \cite{janos:absorp} the authors have found edge-state enhanced transport along the cut between the source and the absorption center in a two-dimensional quantum walk. For continuous-time quantum walks the effects of trapping, scaling and percolation on transport have been discussed in \cite{Muelken:1,Muelken:2,Muelken:3,Darazs:fractal}.

Discrete time quantum walks represent nontrivial dynamics of two
degrees of freedom given by the coin and the position. Due to this
they offer a formidable playground to test the influence of noise,
external perturbations or loss of control over the system
representing all possible aspects of open system dynamics. By
considering walks on changing graph structure a natural link to
percolation theory was established and it is of interest to study
alternation of transport efficiency due to percolation.

We carry out one of the simplest nontrivial percolation assisted
transports. Under transport we understand the gradual leak of the initially localized excitation at a preassigned position called sink. In the present paper we study the transport of the excitation to the sink on a ring graph. The ring graph was chosen as
it is highly symmetric and for even number of sites we can study the
transport across the structure through two equal segments assuming
we choose as source and drain two opposite positions. The ring
structure -- due to its high symmetry -- allows to study the ideal
(non-percolated) dynamics for a large class of quantum walks in a closed form and identify the effect of percolation in quite a transparent way.

We focus on two different scenarios. Namely, the propagation of excitation is described either by a two-state or a three-state (lazy) discrete-time quantum walk. The two-state walk model
where the excitation has to leave its present position was to some
extent analyzed in \cite{konno:absorp,bach:absorp}. The excitation
is fully transferred to the sink independent of the initial
conditions. The rate of the transport depends on the size of the
ring and the coin operator which determines the spreading of the
excitation's wave-packet. This case will serve as the reference for
the more involved lazy walk. When we consider the transport
described by a lazy quantum walk where the particle is
allowed to stay at its actual position, the situation becomes more
interesting. Indeed, certain lazy quantum walks are able to
trap part of the wave-function in the vicinity of the origin
\cite{inui:grover1,inui:grover2}, i.e. the probability of finding
the excitation at finite positions does not vanish in the limit of
infinite number of steps. This feature crucially depends on the
choice of the coin operator \cite{stef:cont:def,stef:3state}. The
consequence of the trapping effect is that the transport of
excitation in the lazy quantum walk is not fully efficient.
Nevertheless, we show that dynamical percolations
\cite{grimmett:perc,kendon:perc} of the ring can eliminate the
trapping effect and improve the transport efficiency.

The paper is organized as follows: In Section ~\ref{sec:1} we
describe the model of excitation transport based on a discrete-time
quantum walk. The two-state walk model is analyzed in
Section~\ref{sec:2}. Section~\ref{sec:3} is devoted to the
lazy walk model. The effect of dynamical percolations on the
transport efficiency is discussed in Section~\ref{sec:4}. We
conclude and present an outlook in Section~\ref{sec:5}. More technical details of calculating the survival probability for a two-state walk are left for \ref{app:2state}.

\section{Transport of excitation in discrete-time quantum walk}
\label{sec:1}

In this Section we formally describe our model. We consider the
transport of excitation to the sink on a finite-size ring described by
a discrete-time quantum walk. The vertices of the ring have labels
from $-N+1$ to $N$. The sink, which takes the excitation away from
the ring, is located at the vertex $N$. Note that this configuration
is equivalent to a finite line from $-N$ to $N$ with sinks at both
ends. The Hilbert space of the quantum walk has the tensor product
structure
$$
{\cal H} = {\cal H}_P\otimes{\cal H}_C,
$$
where ${\cal H}_P$ is the position space spanned by the vectors
$|m\rangle$, with $m=-N+1,\ldots N$, corresponding to the excitation
being at the vertex $m$. By ${\cal H}_C$ we denote the coin space
which describes the internal degree of freedom of the excitation.
The excitation enters the ring exactly opposite the sink with some coin
state $|\psi_C\rangle\in{\cal H}_C$, i.e. the initial state is
$$
|\psi_{in}\rangle = |0\rangle|\psi_C\rangle.
$$
A single time-step consists of a quantum walk evolution $\hat U$
$$
\hat U = \hat S\cdot(\hat I_P\otimes \hat C),
$$
where the step operator $\hat S$ and the coin operator $\hat C$ will be specified later. This is followed by a projection
$$
\hat \pi = \left(\hat I_P-|N\rangle\langle N|\right)\otimes \hat I_C,
$$
corresponding to the effect of the sink. Hence, the complete time
evolution is not unitary and the state of the excitation after $t$
steps is described by vector
$$
|\psi(t)\rangle = \left(\hat \pi\cdot \hat U\right)^t|\psi_{in}\rangle,
$$
with norm generally less than unity.

In the following we analyze the properties of the survival
probability ${\cal P}(t)$, i.e. the probability that the excitation
remains on the ring until time $t$, which is given by
$$
{\cal P}(t) = \langle\psi(t)|\psi(t)\rangle,
$$
and the asymptotic transport efficiency $\eta$ which we define as
$$
\eta = 1 - \lim\limits_{t\rightarrow\infty}{\cal P}(t).
$$
In Section~\ref{sec:2} we consider the two-state quantum walk model,
while in Section~\ref{sec:3} we focus on the lazy model.
Before we turn to these models we first derive an asymptotic estimate
of the survival probability ${\cal P}(t)$. We begin with the
estimate
$$
{\cal P}(t)  \leq \left\|\left(\hat\pi\cdot\hat U\right)^t\right\|^2 = \exp{\left(2t\ln{\left(\left\|\left(\hat\pi\cdot\hat U\right)^t\right\|^\frac{1}{t}\right)}\right)}.
$$
For large $t$ the argument of the logarithm can be estimated according to
$$
\left\|\left(\hat\pi\cdot\hat U\right)^t\right\|^\frac{1}{t} \approx |\lambda_l|,
$$
where $\lambda_l$ is the leading eigenvalue of $\hat\pi\cdot\hat U$, i.e. the largest eigenvalue in absolute value.
When $|\lambda_l|$ is close to unity (but less than one) we make the
first-order Taylor expansion of the logarithm
$$
\ln{\left(\left\|\left(\hat\pi\cdot\hat U\right)^t\right\|^\frac{1}{t}\right)} \approx \ln{|\lambda_l|} \approx -(1-|\lambda_l|).
$$
Hence, when the absolute value of the leading eigenvalue $\lambda_l$ of $\hat\pi\cdot\hat
U$ is smaller then one we find that the survival probability behaves
in the asymptotic limit as an exponential
\begin{equation}
\label{2state:asymp}
{\cal P}(t) \sim e^{-\gamma t},
\end{equation}
where the decay rate $\gamma$ reads
\begin{equation}
\label{2state:decay}
\gamma = 2(1-\left|\lambda_{l}\right|).
\end{equation}
In such a case, the asymptotic transport efficiency $\eta$ is unity. However,
when $|\lambda_l|=1$ the survival probability does not vanish and the
transport is not efficient. We will see in Section~\ref{sec:3} that
such situation occurs in certain lazy quantum walks.

\section{Two-state walk model}
\label{sec:2}

Let us begin with the two-state walk model, i.e. the particle has to
move in each time step from the vertex $m$ to the nearest neighbours
$m-1$ or $m+1$. The coin space ${\cal H}_C$ is two-dimensional, we
denote the basis vectors corresponding to the steps to the left and
right as $|L\rangle$ and $|R\rangle$. The step operator of the
two-state quantum walk is then given by
\begin{equation}
\label{ste:2state}
{\hat S^{(2)}} = \sum_{m=-N+1}^{N} |m-1\rangle\langle m|\otimes|L\rangle\langle L| + |m+1\rangle\langle m|\otimes|R\rangle\langle R|,
\end{equation}
where we consider periodic boundary condition $N\equiv -N$. As for
the coin operator $\hat C$, for simplicity we consider a
one-parameter family which is in the standard basis of the coin
space given by the matrix
\begin{equation}
C^{(2)} = \left(
      \begin{array}{cc}
        \rho & \sqrt{1-\rho^2} \\
        \sqrt{1-\rho^2} & -\rho \\
      \end{array}
    \right), \quad \rho\in (0,1).
\label{coin:2state}
\end{equation}
Nevertheless, this choice in fact covers all $U(2)$ matrices due to
unitary equivalence which was recently found in
\cite{unitary:equivalence}. The coin parameter $\rho$ determines the
rate at which the excitation spreads through the ring \cite{kempf}.
The choice of $\rho=\frac{1}{\sqrt{2}}$ corresponds to the familiar
and extensively studied case of the Hadamard walk.

\begin{figure}[h]
\begin{center}
\includegraphics[width=0.48\textwidth]{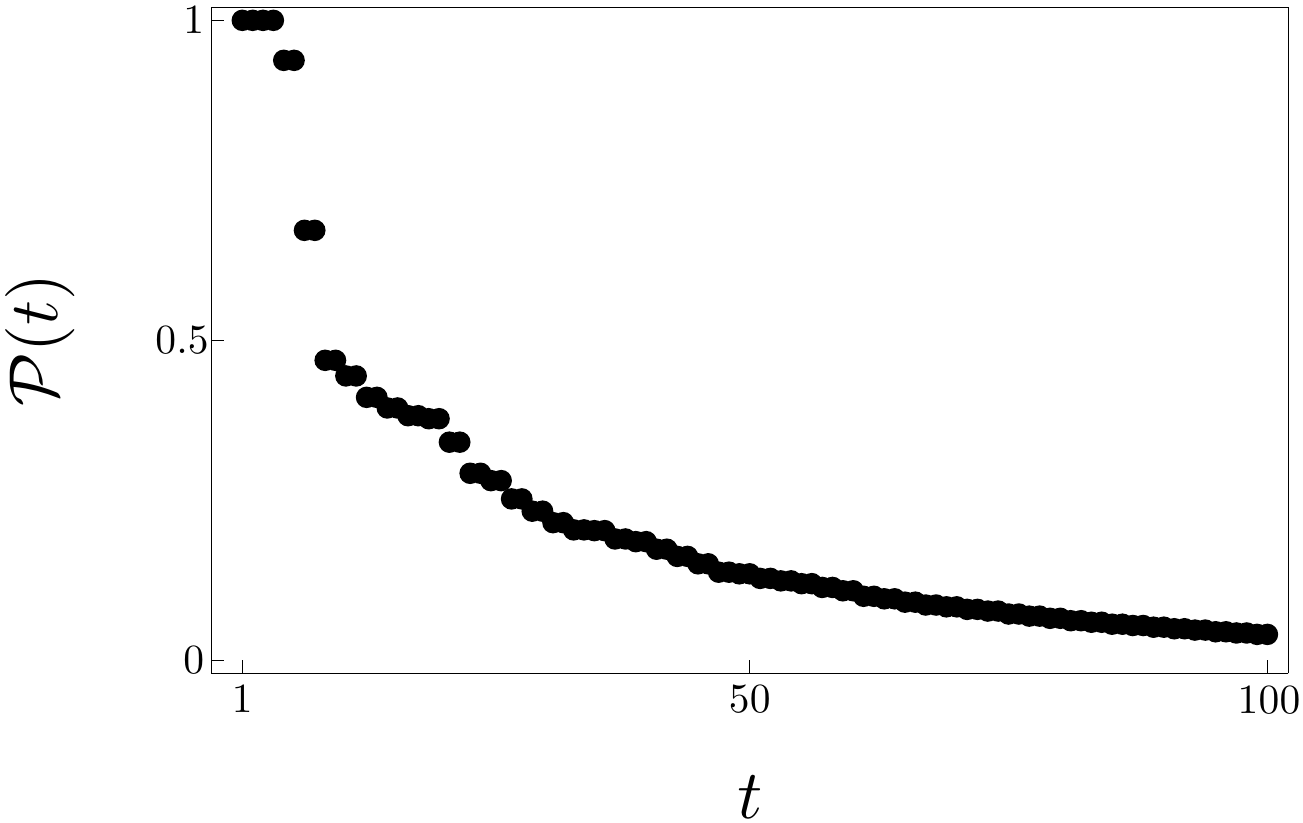}\hfill
\includegraphics[width=0.48\textwidth]{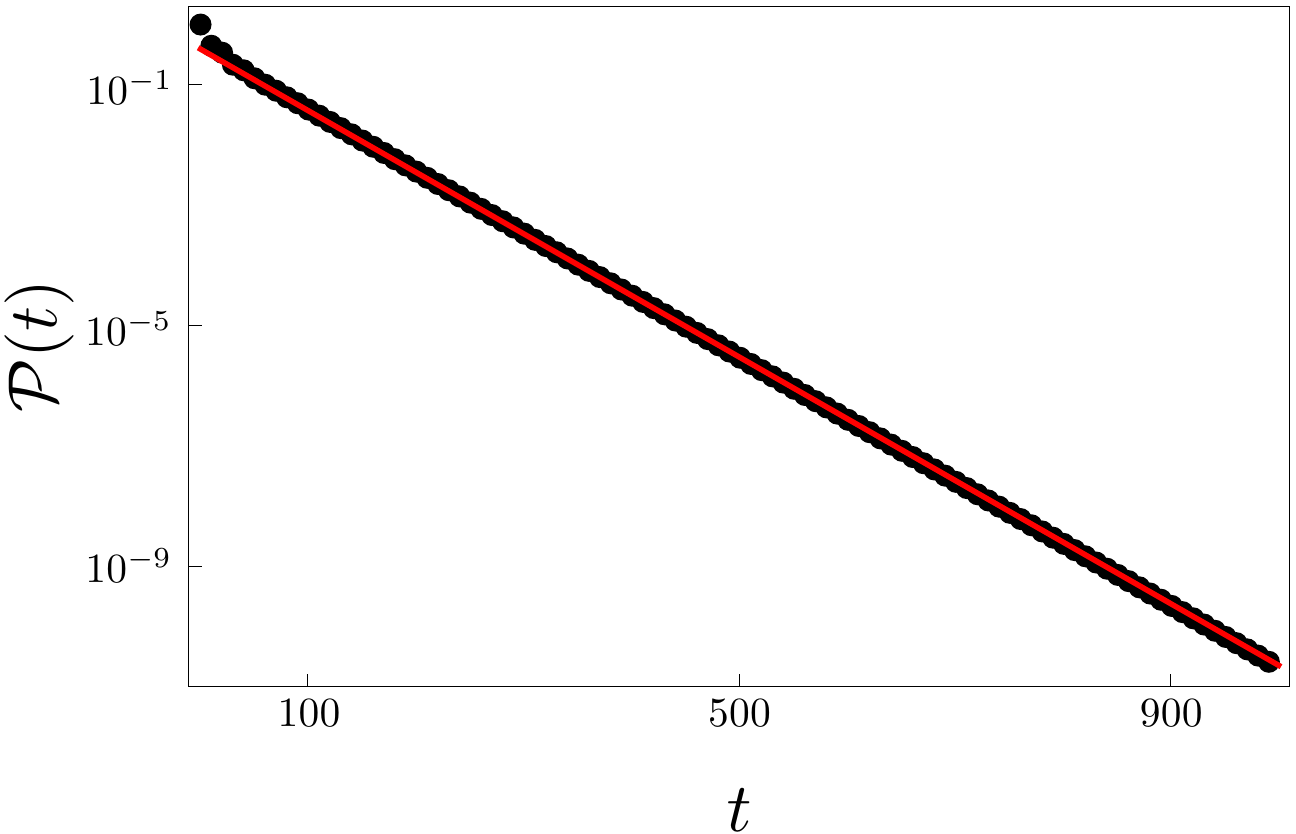}
\end{center}
\caption{On the left we display the survival probability ${\cal
P}(t)$ for the Hadamard walk (i.e. $\rho=\frac{1}{\sqrt{2}}$). Due
to the symmetries of the model the survival probability ${\cal
P}(t)$ is independent of the initial coin state. On the right we
show the asymptotic behavior of the survival probability ${\cal
P}(t)$ for the Hadamard walk. The log-scale reveals that the
survival probability decays exponentially. The red curve is given by
(\ref{2state:asymp}). The size of the ring is given by $N=5$, i.e.
10 vertices.}
\label{fig:surv:had}
\end{figure}

In Figure~\ref{fig:surv:had} we present the numerical simulation of
the survival probability ${\cal P}(t)$ for the Hadamard walk
($\rho=\frac{1}{\sqrt{2}}$) on the ring consisting of 10 vertices,
i.e. $N=5$. The left plot shows the survival probability for the
first 100 steps. Due to the symmetries of the model we consider,
i.e. the excitation enters the ring exactly opposite the sink,
the survival probability is exactly the same for all initial coin
states $|\psi_C\rangle$. This follows from the results of
\cite{konno:absorp}, as we show in \ref{app:2state}. The right plot
displays the survival probability on a longer time scale of 1000
steps. To unravel the asymptotic behavior of ${\cal P}(t)$ we use
logarithmic scale on the $y$ axis. The plot indicates that the
survival probability decays exponentially (\ref{2state:asymp}) with
the decay rate (\ref{2state:decay}) determined by the leading
eigenvalue of $\hat\pi\cdot \hat U$.

\begin{figure}[h]
\begin{center}
\includegraphics[width=0.6\textwidth]{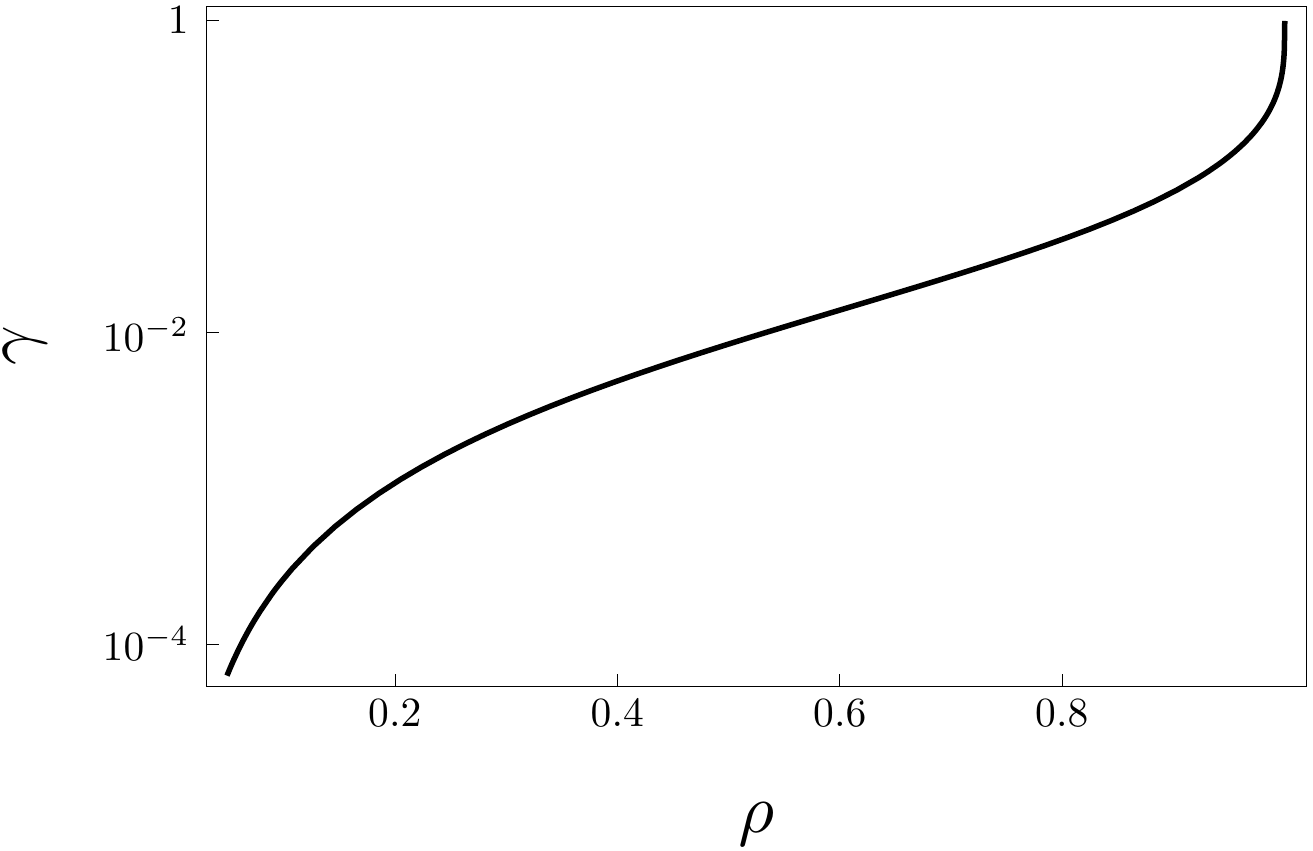}
\end{center}
\caption{The decay rate $\gamma$ as a function of the coin parameter $\rho$. The size of the ring is given by $N=5$.}
\label{fig:decay:2state}
\end{figure}

The decay rate $\gamma$ depends crucially both on the coin parameter
$\rho$ and the size of the ring $N$. We plot the decay rate $\gamma$
as a function of the coin parameter $\rho$ in
Figure~\ref{fig:decay:2state}. For $\rho$ approaching zero the decay
rate vanishes. Indeed, for $\rho=0$ the coin operator
(\ref{coin:2state}) turns into a permutation matrix. In such a case,
the excitation never leaves the vertices -1, 0 and 1, i.e. it never
reaches the sink and the asymptotic transport efficiency $\eta$ is
zero, except for $N=1$ when the ring consists only of the source and the sink. On the other hand, for $\rho$
approaching unity the decay rate increases rapidly. In the limiting
case of $\rho=1$ the coin operator (\ref{coin:2state}) reduces to an
identity with an additional phase shift of $\pi$ on the $|R\rangle$
state. Hence, the excitation is fully transported into the sink
after $N$ steps and the survival probability
is a step function, with ${\cal P}(t)=1$ for $t<N$ and ${\cal
P}(t)=0$ for $t\geq N$.

\begin{figure}[h]
\begin{center}
\includegraphics[width=0.48\textwidth]{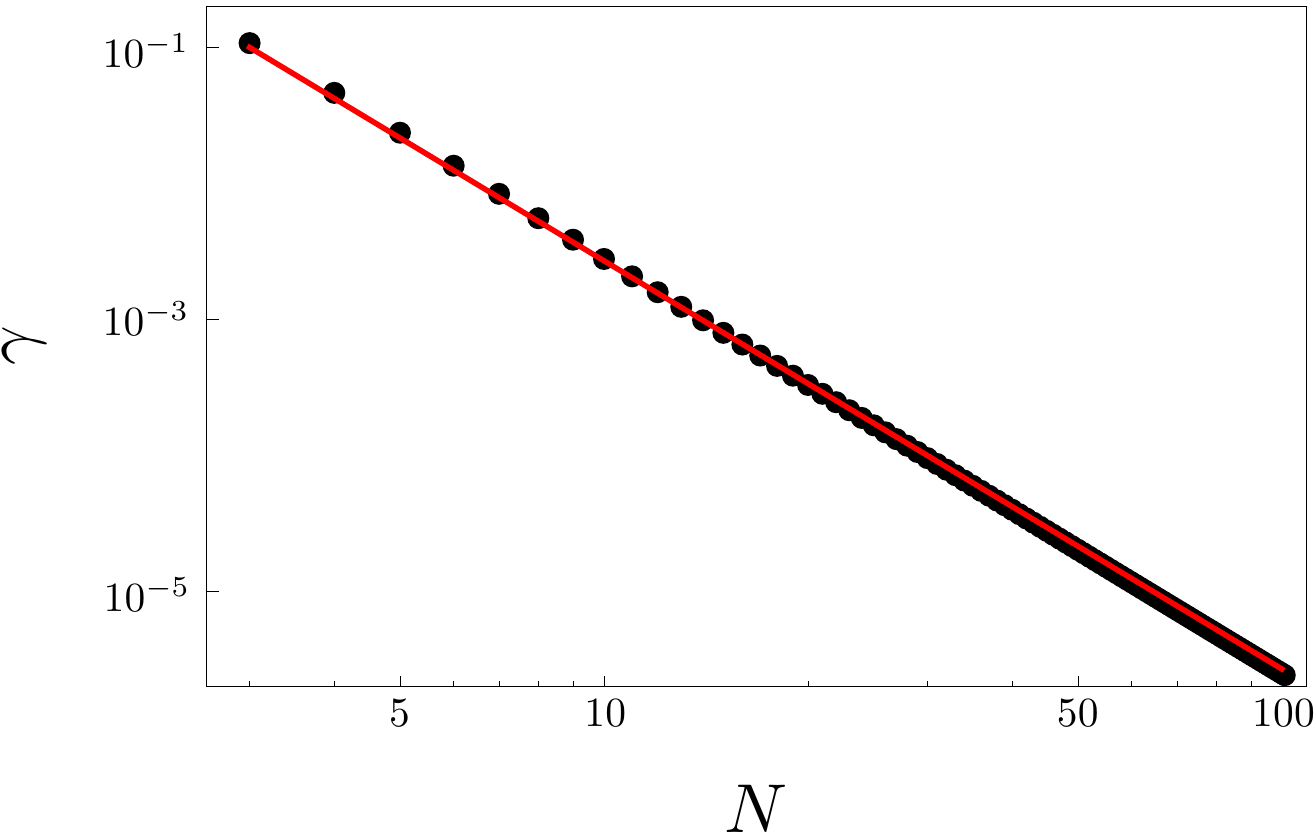}\hfill
\includegraphics[width=0.48\textwidth]{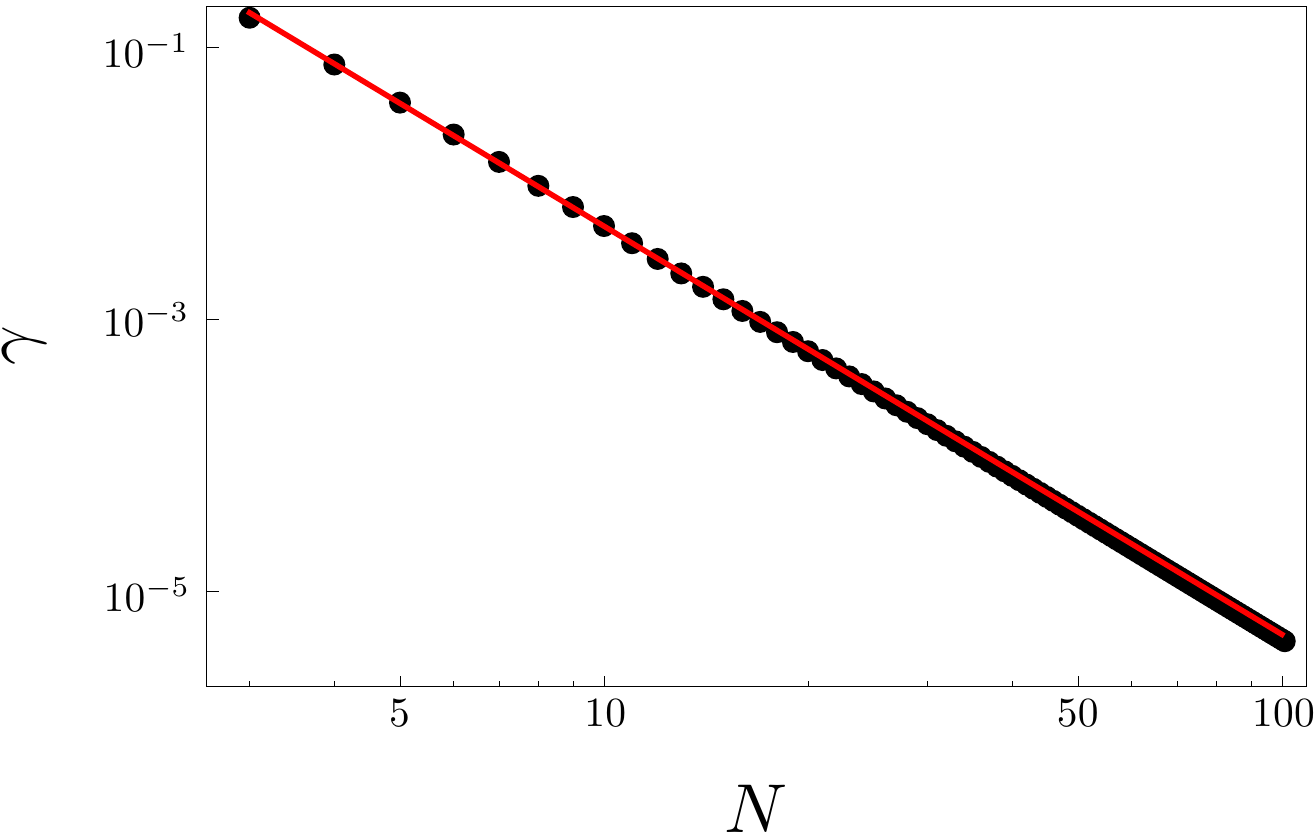}
\end{center}
\caption{The decay rate $\gamma$ as a function of the size of the ring $N$. Left plot shows the choice $\rho=\frac{1}{\sqrt{2}}$ corresponding to the Hadamard walk, while in the right plot we have chosen $\rho=0.8$. The plots indicate that the decay rate rate obeys a power law (\ref{2state:decay:N}). This behaviour is independent of the coin parameter $\rho$.}
\label{fig:decay:2state:N}
\end{figure}

In Figure~\ref{fig:decay:2state:N} we consider the decay rate
$\gamma$ as a function of the size of the ring $N$. For the plot on
the left we have chosen the coin parameter $\rho=\frac{1}{\sqrt{2}}$
corresponding to the Hadamard walk, while in the right plot we have
considered $\rho=0.8$. The log-log scale reveals that the decay rate
obeys a power law
\begin{equation}
\gamma\sim N^{-3},
\label{2state:decay:N}
\end{equation}
independent of the coin parameter $\rho$.

\begin{figure}[h]
\begin{center}
\includegraphics[width=0.48\textwidth]{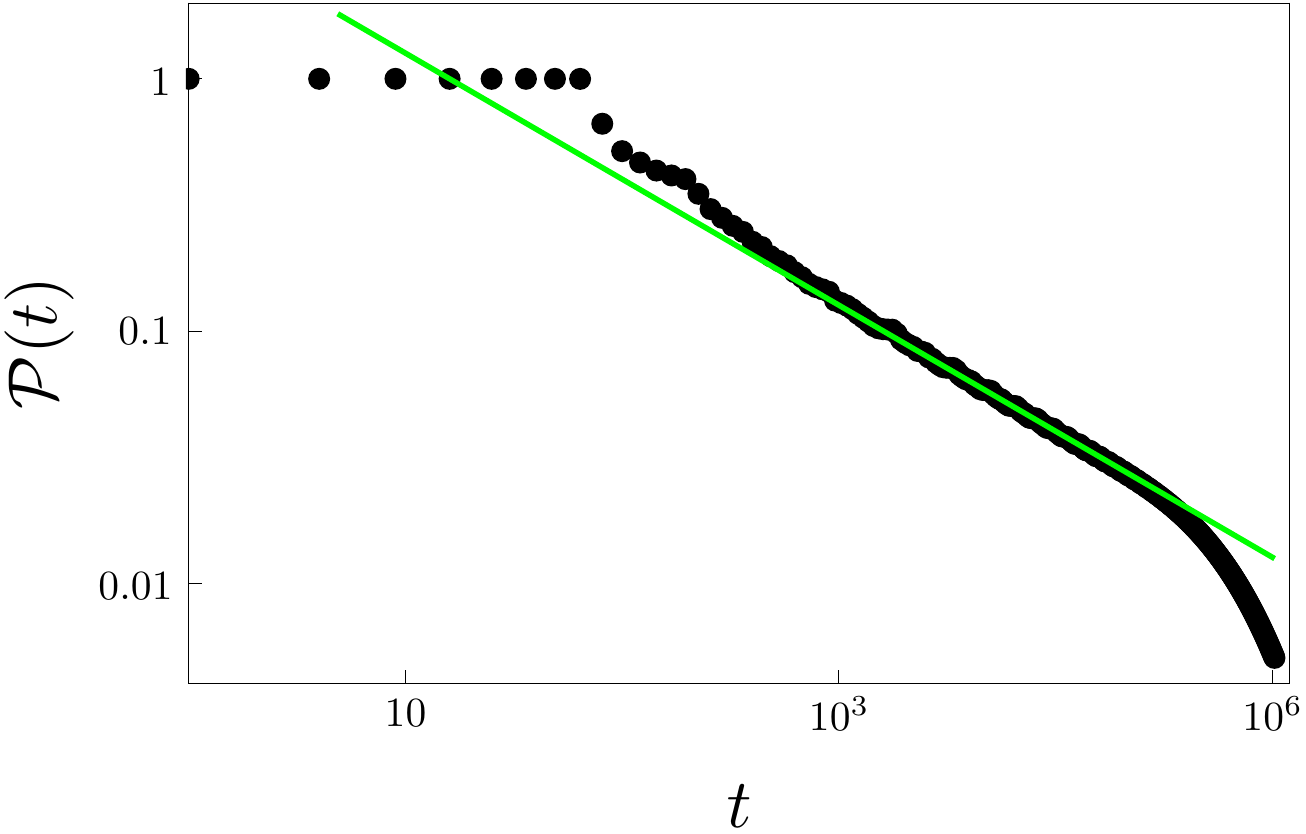}\hfill
\includegraphics[width=0.48\textwidth]{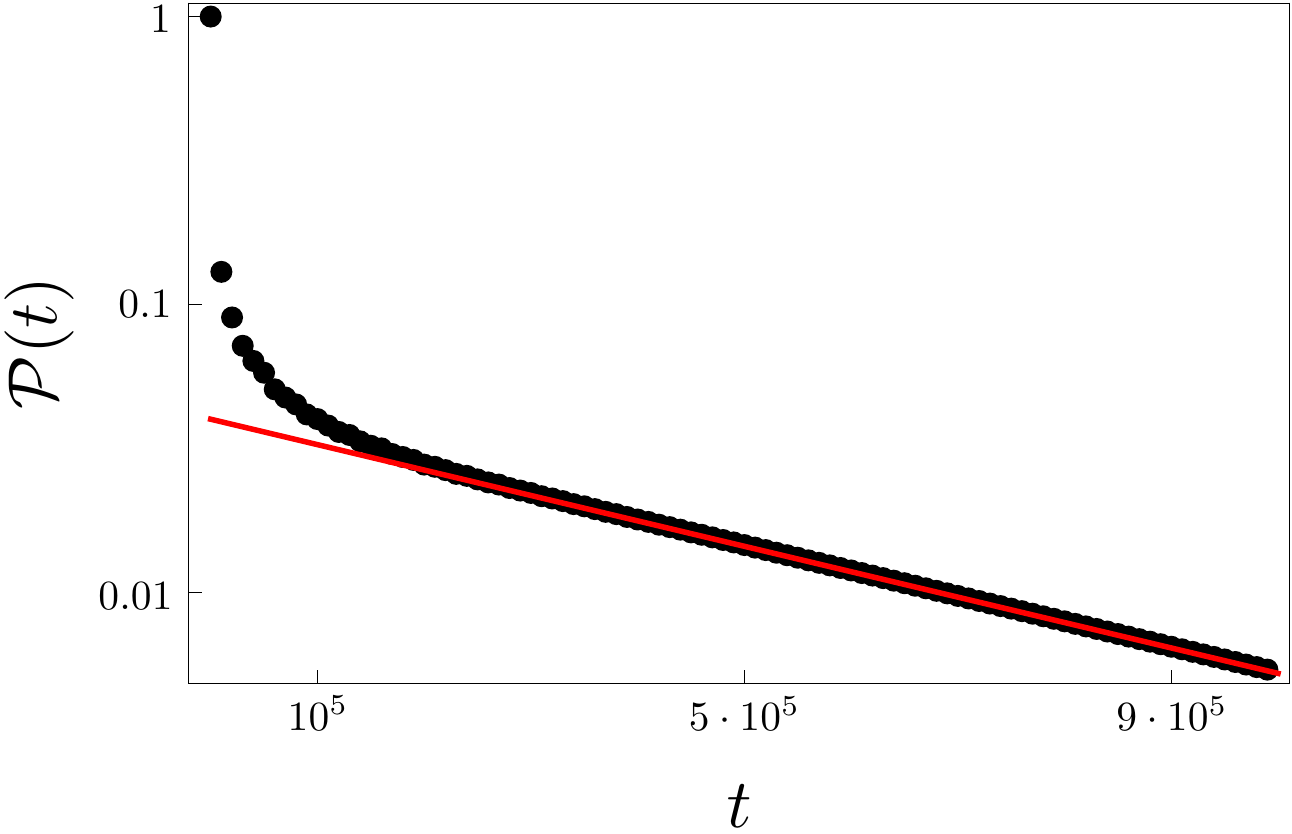}
\end{center}
\caption{Survival probability for the Hadamard walk on a ring with
100 vertices, i.e. $N=50$. On the left we display ${\cal P}(t)$ on a
log-log scale. The green line corresponds to the power law behavior
of (\ref{2state:plaw}). We observe that it fits well in the
intermediate regime, however, it deviates on the long time scale.
Indeed, the plot on the right, where we display the survival
probability on a log-scale, shows that in the asymptotic regime the
behavior of ${\cal P}(t)$ follows the exponential
(\ref{2state:asymp}) depicted by the red line.}
\label{fig:psurv:dev}
\end{figure}

We point out that for large $N$ the decay rate $\gamma$ is very
small. Hence, it takes a considerable number of steps for the
exponential behaviour of the survival probability
(\ref{2state:asymp}) to set in. As was established and explained at length in Section 4.6 of \cite{bach:absorp}, repeated reflections from the sink (or, in fact, from the adjacent points) slow the decay of the survival probability in the intermediate regime to a power law
\begin{equation}
\label{2state:plaw}
{\cal P}(t) \sim t^{-\frac{1}{2}}.
\end{equation}
Nevertheless, after an order of $N^3$ steps the survival probability
begins to deviate from the power law (\ref{2state:plaw}) and tends
to follow the exponential dependence on the number of steps
(\ref{2state:decay}). We illustrate this property in
Figure~\ref{fig:psurv:dev}.

To summarize this section, for the two-state model the survival
probability ${\cal P}(t)$ is independent of the initial coin state
and decays exponentially. Hence, the asymptotic transport efficiency
$\eta$ is unity.

\section{Lazy walk model}
\label{sec:3}

We now turn to the lazy walk model. Let us denote the basis
coin states corresponding to the step to the left, stay, and the
step to the right as $|L\rangle$, $|S\rangle$ and $|R\rangle$. The
step operator of a lazy quantum walk on a ring is then given
by the following extension of the step operator for a two-state walk
(\ref{ste:2state})
$$
{\hat S^{(3)}} = {\hat S^{(2)}} + \sum_{m=-N+1}^{N} |m\rangle\langle m|\otimes|S\rangle\langle S|.
$$
Next, we choose the coin operator which exhibits the trapping
effect. The complete set of such coins for a lazy walk was
determined in \cite{stef:3state}. For simplicity, we consider a
two-parameter set of coins which are in the standard basis of the
coin space $\{|L\rangle, |S\rangle, |R\rangle\}$ given by the matrix
\begin{equation}
\label{coin:3state}
C^{(3)} = \left(
\begin{array}{ccc}
 -\rho^2 & \rho \sqrt{2-2 \rho^2} & e^{-i \alpha} \left(1 - \rho^2\right) \\
 \rho \sqrt{2-2 \rho^2} & 2 \rho^2-1 & e^{-i \alpha} \rho \sqrt{2-2 \rho^2} \\
 e^{i \alpha} \left(1-\rho^2\right) & e^{i \alpha} \rho \sqrt{2-2 \rho^2} & -\rho^2 \\
\end{array}
\right)
\end{equation}
where $\rho$ ranges from zero to one and the phase $\alpha$ from 0
to $2\pi$. The parameter $\rho$ has the same interpretation as for
the two-state walk of Section~\ref{sec:2}. The phase $\alpha$ will
not play a role in this Section, however, it will be crucial when we
consider percolations in Section~\ref{sec:4}.

The trapping effect arises when the evolution operator of the
quantum walk has a highly degenerate eigenvalue and the
corresponding eigenstates are spatially localized
\cite{inui:grover1,inui:grover2}. One can show by direct calculation
that this is the case for the lazy walk on ring with the coin
(\ref{coin:3state}). Indeed, the evolution operator $\hat U$ has an
eigenvalue $\lambda=1$ with $2N$-fold degeneracy and the
corresponding eigenvectors (linearly independent but overlapping)
read
\begin{eqnarray}
\nonumber |s_n\rangle & = & |n\rangle\left(
                \sqrt{1-\rho^2}|L\rangle + \frac{\rho}{\sqrt{2}}|S\rangle\right) + \\
& & +|n+1\rangle\left(\frac{\rho}{\sqrt{2}}|S\rangle + e^{i\alpha}\sqrt{1-\rho^2}|R\rangle
     \right),
\label{stat:state}
\end{eqnarray}
where $n$ ranges from $-N+1$ to $N$. Notice that only two of these
vectors, namely $|s_{N-1}\rangle$ and $|s_{N}\rangle$, have support
on the vertex $N$ where the sink is located. Hence, the vectors
$|s_n\rangle$ with $n\in\{-N+1, \ldots, N-2\}$ are not affected by
the sink and they are eigenvectors of $\hat\pi\cdot \hat U$ with
eigenvalue one. Consequently, the trapping effect remains even in
the presence of the sink, the survival probability has a
non-vanishing limit and the excitation transport is not fully
efficient.

Let us now evaluate the transport efficiency $\eta$. Using the
Gram-Schmidt procedure one can form an orthonormal basis in the
degenerate subspace from the eigenstates (\ref{stat:state}). We
denote the basis vectors by $|\phi_n\rangle$. The probability of
trapping the excitation on the vertex $m$, i.e. the probability of
finding the excitation at position $m$ in the limit of infinite
number of steps, is obtained from
\begin{equation}
\label{p:trap}
p_T(m) = \sum\limits_{i=L,S,R} \left|\langle m|\langle i|\left(\sum\limits_n |\phi_n\rangle\langle\phi_n|\right) |\psi_{in}\rangle\right|^2 .
\end{equation}
The limiting value of the survival probability is then given by summing the trapping probabilities over all vertices of the ring excluding the sink
$$
\lim\limits_{t\rightarrow\infty}{\cal P}(t) = \sum\limits_{m= - N+1}^{N-1} p_T(m),
$$
which can be simplified into
$$
\lim\limits_{t\rightarrow\infty}{\cal P}(t) = \sum_{n=-N+1}^{N-2}|\langle\psi_{in}|\phi_n\rangle|^2.
$$
Hence, the asymptotic transport efficiency reads
$$
\eta = 1 - \sum_{n=-N+1}^{N-2}|\langle\psi_{in}|\phi_n\rangle|^2.
$$
The evaluation of $\eta$ is readily done for small $N$. We present the results for $N=2,\ldots ,5$ in Table~\ref{table:eta}.

\begin{table}[h]
\begin{center}
\begin{tabular}{|c|c|}
  \hline
  $N$ & $\eta$ \\
 \hline
 & \\
  2 & $1 - \frac{2(1-\rho^2)}{4-3\rho^2}|h_2|^2 - \frac{2}{4-\rho^2}|h_+|^2$ \\
  & \\
  \hline
  & \\
  3 & $1 - 4(2-\rho^2)\left( \frac{(1-\rho^2)|h_2|^2}{16-20\rho^2+5\rho^4} + \frac{|h_+|^2}{16-12\rho^2+\rho^4}\right)$ \\
  & \\
  \hline
   &  \\
   4 & $1 - 2(16-16\rho^2+3\rho^4)\left(\frac{(1-\rho^2)|h_2|^2}{64-7\rho^2(\rho^2-4)^2} + \frac{|h_+|^2}{64-\rho^2(\rho^4-24\rho^2+80)}\right)$ \\
   &  \\
   \hline
   &  \\
   5 & $1- 8 \left(2-\rho^2\right) \left(\rho^4-8 \rho^2+8\right) \left(\frac{\left(1-\rho^2\right) \left|h_2\right| ^2}{\left(3 \rho^2-4\right) \left(3 \rho^6-36 \rho^4+96 \rho^2-64\right)} + \frac{\left|h_+\right| ^2}{\rho^8-40 \rho^6+240 \rho^4-448 \rho^2+256}\right)$ \\
& \\
  \hline
\end{tabular}
\caption{Asymptotic transport efficiency $\eta$ for small rings up to $N=5$.}
\label{table:eta}
\end{center}
\end{table}

In order to reduce the complexity of the formulas we have expressed
the initial coin state $|\psi_C\rangle$ in terms of a more suitable
basis of the coin space. Following \cite{stef:eigen} we have chosen
the basis formed by the eigenvectors of the coin operator
(\ref{coin:3state})
\begin{eqnarray}
\label{eigenbasis}
\nonumber |\sigma^+\rangle & = & \sqrt{\frac{1 - \rho^2}{2}}|L\rangle +\rho|S\rangle + \sqrt{\frac{1 - \rho^2}{2}}e^{i\alpha}|R\rangle, \\
\nonumber |\sigma_1^-\rangle & = & \frac{\rho}{\sqrt{2}}|L\rangle -\sqrt{1-\rho^2}|S\rangle + \frac{\rho}{\sqrt{2}}e^{i\alpha}|R\rangle, \\
|\sigma_2^-\rangle & = & \frac{1}{\sqrt{2}}(|L\rangle - e^{i\alpha}|R\rangle).
\end{eqnarray}
The initial coin state is in the eigenbasis decomposed according to
$$
|\psi_C\rangle = h_+|\sigma^+\rangle + h_1 |\sigma_1^-\rangle + h_2 |\sigma_2^-\rangle.
$$
There are several advantages of using the basis (\ref{eigenbasis}).
First, $\eta$ is independent of the amplitude $h_1$, as can be seen
from Table~\ref{table:eta}. Indeed, for $h_1=1$ and $h_+=h_2=0$ the
trapping effect vanishes \cite{stef:eigen}. Hence, the initial coin
state $|\sigma_1^-\rangle$ is the only one for which the transport
efficiency $\eta$ is unity. Next, $\eta$ does not depend on the
phase $\alpha$ which was absorbed into the definition of the
eigenbasis (\ref{eigenbasis}). Thus, the coins with different values
of $\alpha$ are equivalent\footnote{This is no longer true when we
consider percolations of the ring, as we will show in the following
Section.}. Finally, the amplitudes $h_+$ and $h_2$ enter the formula
for the transport efficiency $\eta$ only as probabilities $|h_+|^2$
and $|h_2|^2$ of finding the particle initially in the coin state
$|\sigma^+\rangle$ or $|\sigma_2^-\rangle$. Hence, the
efficiency of transfer is given by incoherent contributions from the
two relevant basis states. It is then straightforward to show that the worst transport efficiency arises when the initial coin state is chosen as
$|\sigma^+\rangle$.

\begin{figure}[h]
\begin{center}
\includegraphics[width=0.48\textwidth]{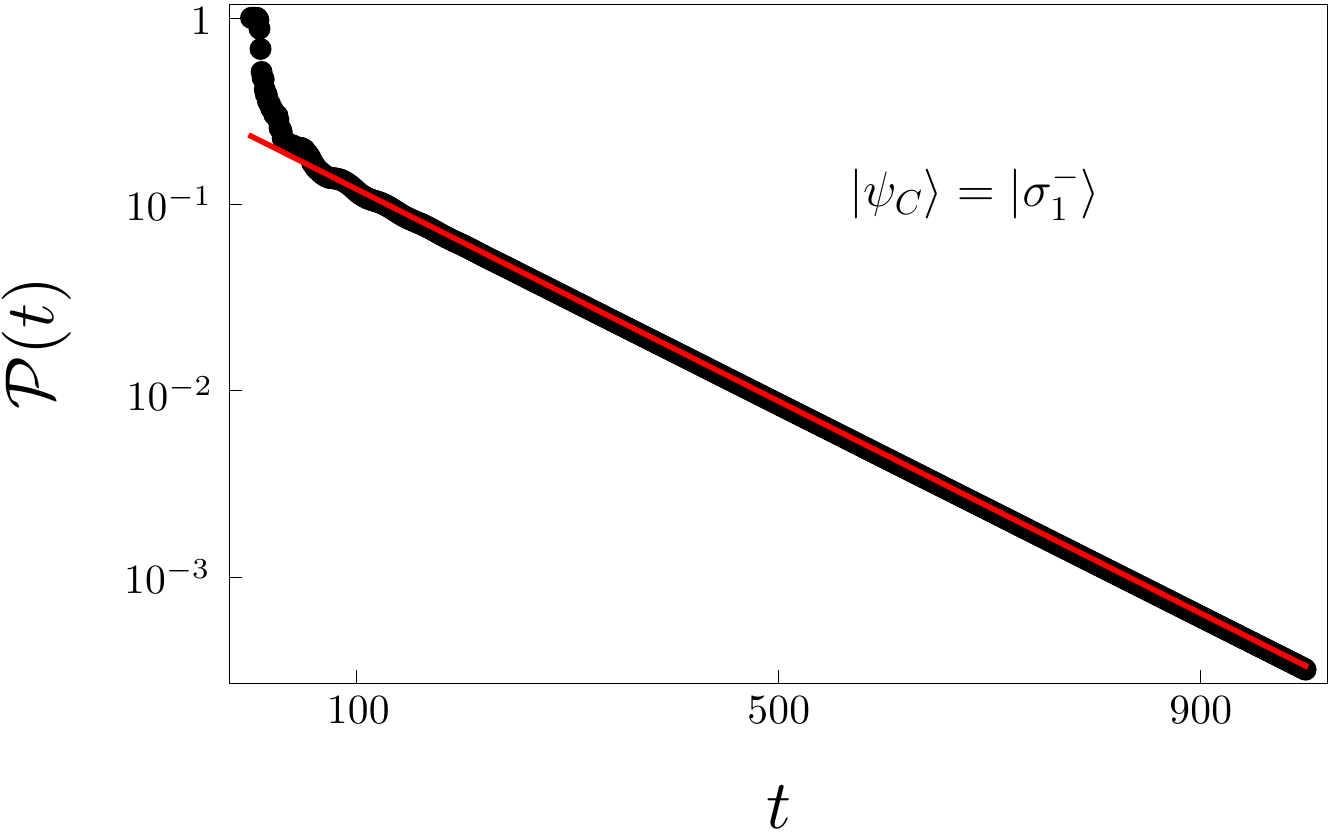}\hfill
\includegraphics[width=0.48\textwidth]{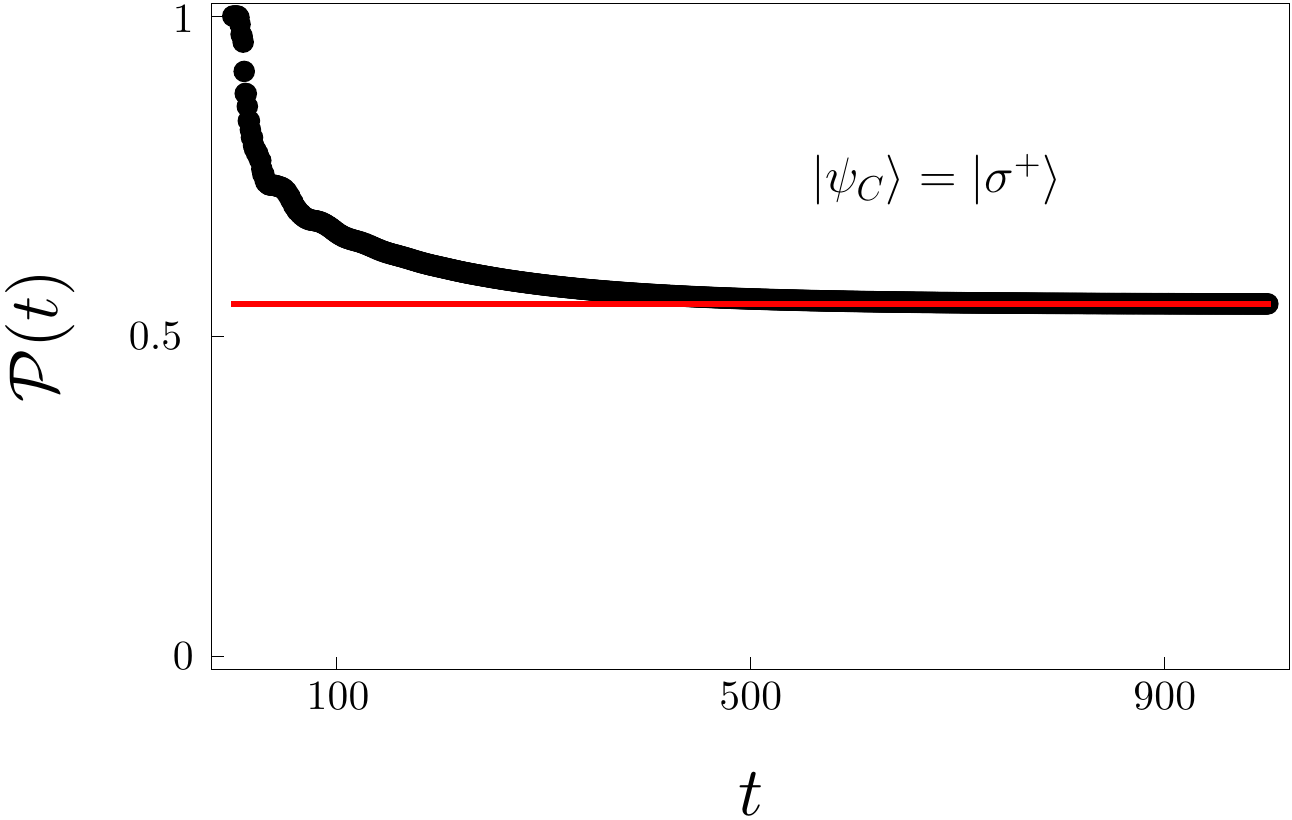}
\end{center}
\caption{Survival probability for the Grover walk on a ring with 10
vertices, i.e. $N=5$ in dependence on the number of steps. On the
left we have chosen the initial coin state as $|\sigma_1^-\rangle$
for which the trapping effect disappears. In such a case the
survival probability ${\cal P}(t)$ vanishes exponentially. This is
highlighted by the log-scale. The red line corresponds to the decay
rate $\gamma$ determined by (\ref{3state:decay}). The right plot
shows the survival probability when the initial coin state is
$|\sigma^+\rangle$ for which the trapping effect is the strongest.
The survival probability approaches the red line given by
$\sum\limits_{m= - N+1}^{N-1} p_T(m)\approx 0.55$.}
\label{fig:3state}
\end{figure}

To illustrate our results we display in Figure~\ref{fig:3state} the
survival probability for the Grover walk (i.e.
$\rho=\frac{1}{\sqrt{3}}$ and $\alpha = 0$), when the initial coin
state is chosen as $|\sigma_1^-\rangle$ (left plot) or
$|\sigma^+\rangle$ (right plot). For $|\sigma_1^-\rangle$ the
trapping effect disappears and the survival probability decays
exponentially (\ref{2state:asymp}), similarly to the two-state walk
of Section~\ref{sec:2}. The decay rate $\gamma$ can be estimated
using the sub-leading eigenvalue $\lambda_{sl}$ of $\hat\pi\cdot\hat
U$ according to
\begin{equation}
\label{3state:decay}
\gamma = 2(1-|\lambda_{sl}|).
\end{equation}
Nevertheless, for all other initial coin states the trapping effect
results in non-vanishing limit of the survival probability. For
$|\sigma^+\rangle$ the trapping effect is the strongest. The right
plot indicates that the survival probability does not drop below the
value $\sum\limits_{m= - N+1}^{N-1} p_T(m)\approx 0.55$, which is
depicted by the red line.

For larger rings the Gram-Schmidt procedure becomes tedious and,
moreover, the resulting formula for the transport efficiency $\eta$
is rather lengthy. Nevertheless, we can estimate the transport
efficiency following the analysis of the walk on an infinite line.
We approximate the trapping probability at position $m$
(\ref{p:trap}) using the results obtained for infinite line in
\cite{stef:eigen} where it was found
\begin{equation}
\label{loc:rho}
p_T(m) = \left\{
                \begin{array}{c}
                  \frac{2-2\rho^2}{\rho^4}Q^{2m} |h_+ + h_2|^2 ,\quad m>0 , \\
                   \\
                   \frac{Q}{\rho^2}\left\{|h_+|^2 + (1-\rho^2)|h_2|^2 \right\}, \quad m = 0.\\
                   \\
                  \frac{2-2\rho^2}{\rho^4}Q^{2|m|} |h_+ - h_2|^2 ,\quad m<0 \\
                \end{array}
              \right.
\end{equation}
Here the quotient $Q$ reads
$$
Q = \frac{2-\rho^2-2\sqrt{1-\rho^2}}{\rho^2}.
$$
Hence, the asymptotic transport efficiency for a ring of size $2N$ can be estimated by
\begin{eqnarray}
\nonumber \eta & \approx 1 - \frac{Q}{\rho^2}  & \left(  \sqrt{1-\rho^2}\left(1-Q^{2(N-1)}\right)\left(|h_2|^2 + |h_+|^2\right) + \right. \\
\nonumber & & \left. +(1-\rho^2)|h_2|^2 + |h_+|^2\frac{}{} \right).
\end{eqnarray}
Since the trapping probability (\ref{loc:rho}) decays very fast
(exponentially) with the distance from the origin, this
approximation is quite good even for small rings. The difference is
most profound for the coin parameter $\rho$ close to one. With
increasing size of the ring the differences become negligible. This
result also confirms that the worst transport efficiency is obtained
for the initial coin state $|\sigma^+\rangle$.

\section{Dynamical percolation of the ring}
\label{sec:4}

In this Section we analyze the effect of dynamical percolation of
the ring on the transport efficiency of the lazy quantum walk. Percolation can be viewed as a special (but realistic) noise source and hence the
problem at hand can be cast under the headline of noise assisted
excitation transfer. Improving  transport by
allowing the edges to break randomly seems to be a bit
counterintuitive at the first sight. However, percolations can in
some situations eliminate the localized eigenstates
(\ref{stat:state}) and thus improve the asymptotic transport efficiency to
unity.

The evolution of the percolated quantum walk can be described within
the framework of random unitary channels
\cite{balint:perc1,balint:perc2}. The density matrix of the
excitation evolves according to the formula
\begin{equation}
\label{ruch}
\hat\rho'(t+1) = \sum_{\mathcal{K}}p_{\mathcal K} \hat U_{\mathcal K}\hat\rho(t) \hat U_{\mathcal K}^\dagger\hat,
\end{equation}
where $\mathcal K$ denotes the possible edge configuration,
$p_{\mathcal K}$ is the probability of the configuration $\mathcal
K$ and $\hat U_{\mathcal K}$ is a quantum walk on a ring with edge
configuration $\mathcal K$. The random unitary channel (\ref{ruch})
is followed by the projection
\begin{equation}
\label{rho:sink}
\hat\rho(t+1) = \hat\pi\hat\rho'(t+1)\hat\pi^\dagger,
\end{equation}
which corresponds to the action of the sink. For simplicity, we
consider that every edge occurs with the same probability $p$
independent of its position. The probability of the edge
configuration ${\mathcal K}$ is then given by
$$
p_{\mathcal K} = p^{|\mathcal K|}(1-p)^{2N-|\mathcal K|},
$$
where $|\mathcal K|$ denotes the size of the set $\mathcal K$, i.e.
the number of edges present in that configuration. The evolution
operator $\hat U_{\mathcal K}$ of the walk on a percolated ring with
edge configuration $\mathcal K$ has the form
$$
\hat U_{\mathcal K} = \hat S_{\mathcal K}\cdot(\hat I_P\otimes \hat C),
$$
where $\hat S_{\mathcal K}$ is the step operator on the percolated
ring. If the edge between $m$ and $m+1$ is broken then the jumps
from $m$ to $m+1$ and from $m+1$ to $m$ cannot occur. Instead, the
coin states corresponding to the jumps undergoes a reflection, i.e.
\begin{equation}
\label{reflection}
|m\rangle|R\rangle \rightarrow |m\rangle|L\rangle,\quad |m+1\rangle|L\rangle \rightarrow |m+1\rangle|R\rangle.
\end{equation}
Hence, the step operator $\hat S_{\mathcal K}$ on the percolated ring is given by
\begin{eqnarray}
\nonumber \hat S_{\mathcal K} & = & \sum\limits_{(m,m+1)\in\mathcal K} |m\rangle\langle m+1|\otimes|L\rangle\langle L| + |m+1\rangle\langle m|\otimes|R\rangle\langle R|+  \\
\nonumber & & + \sum\limits_{(m,m+1)\notin\mathcal K} |m\rangle\langle m|\otimes|L\rangle\langle R| + |m+1\rangle\langle m+1|\otimes|R\rangle\langle L| + \\
\nonumber & & + \sum\limits_{m} |m\rangle\langle m|\otimes|S\rangle\langle S|.
\end{eqnarray}

The evolution of a dynamically percolated quantum walk is rather
involved. Nevertheless, it simplifies considerably in the asymptotic
regime where it is described by the attractors satisfying
$$
\hat U_{\mathcal K}\hat X \hat U_{\mathcal K}^\dagger = \lambda \hat X, \quad \forall {\mathcal K},\quad {\rm with}\  |\lambda|=1 .
$$
Moreover, substantial part of the attractor space is spanned by the so-called p-attractors \cite{balint:perc1,balint:perc2}, which can be constructed from the common eigenstates of all $\hat U_{\mathcal K}$'s. The common eigenstate $|\xi\rangle$ has to satisfy the equations
$$
\hat U_{\mathcal K} |\xi\rangle = \beta |\xi\rangle,
$$
for all possible configurations $\mathcal K$. We search for the common eigenstates in the form
$$
|\xi\rangle = \sum\limits_m |m\rangle|\xi^m\rangle,
$$
where the coin state at position $m$ is given by
$$
|\xi^m\rangle = \xi_L^m|L\rangle + \xi_S^m|S\rangle + \xi_R^m|R\rangle.
$$
Following \cite{balint:perc1,balint:perc2} we find that the amplitudes of the common eigenstate have to fulfill the shift conditions
\begin{equation}
\label{shift:cond}
\xi_L^m = \xi_R^{m+1},\quad \forall m.
\end{equation}
Moreover, the common eigenstates have to fulfill the coin conditions
\begin{equation}
\label{coin:cond}
\hat {\cal R}\hat C|\xi^m\rangle = \beta|\xi^m\rangle, \quad \forall m,
\end{equation}
where $\hat{\cal R}$ is the reflection operator which performs the operation (\ref{reflection}). In the standard basis of the coin space it is given by the matrix
$$
{\cal R} = \left(
      \begin{array}{ccc}
        0 & 0 & 1 \\
        0 & 1 & 0 \\
        1 & 0 & 0 \\
      \end{array}
    \right).
$$
Let us now test when the stationary states of the non-percolated (ideal) walk satisfy the common eigenstates conditions. Form (\ref{stat:state}) it follows that the amplitudes of the stationary state $|s_n\rangle$ are given by
\begin{eqnarray}
\nonumber \xi^m_L & = & \delta_{m,n}\sqrt{1-\rho^2},\\
\nonumber \xi^m_S & = & (\delta_{m,n} + \delta_{m,n+1})\frac{\rho}{\sqrt{2}},\\
\nonumber \xi^m_R & = & \delta_{m,n+1}\sqrt{1-\rho^2}e^{i\alpha}.
\end{eqnarray}
Hence, we find that the shift conditions (\ref{shift:cond}) are
fulfilled only for $\alpha = 0$. One can check that the coin
conditions (\ref{coin:cond}) are also satisfied only in this case.
Hence, for $\alpha=0$ the percolations do not eliminate the
stationary states (\ref{stat:state}) - they remain as common
eigenstates of all $\hat U_{\mathcal K}$'s. Moreover, the stationary
states $|s_n\rangle$ with $n\in\{-N+1, \ldots, N-2\}$ are not
affected by the projection $\hat\pi$ corresponding to the effect of
the sink. Therefore, for $\alpha=0$ the trapping effect is preserved
in the percolated walk and the efficiency of transport to the sink
is not improved, i.e. $\eta$  depends on the initial coin state in
the same way as for the ideal walk. On the other hand, for $\alpha
\neq 0$ the stationary states (\ref{stat:state}) do not satisfy the
common eigenstates conditions and they are sensitive to
percolations. Hence, for $\alpha\neq 0$ dynamical percolations of
the ring eliminate the trapping effect and the transport of
excitation is efficient, i.e. $\eta=1$ for all initial coin states. We
see that percolations nullify the equivalence of coins with
different values of $\alpha$ which holds for ideal walks.

\begin{figure}[h]
\begin{center}
\includegraphics[width=0.48\textwidth]{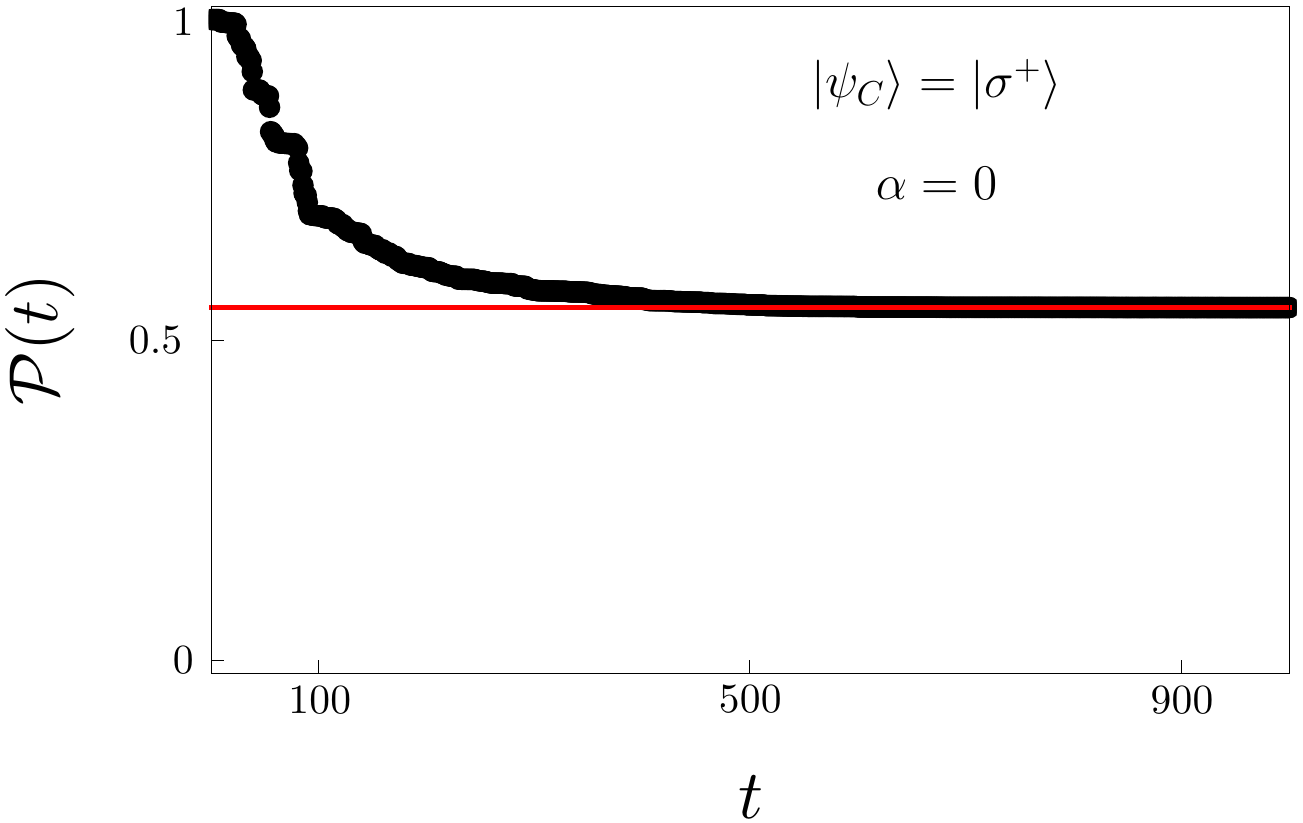}\hfill
\includegraphics[width=0.48\textwidth]{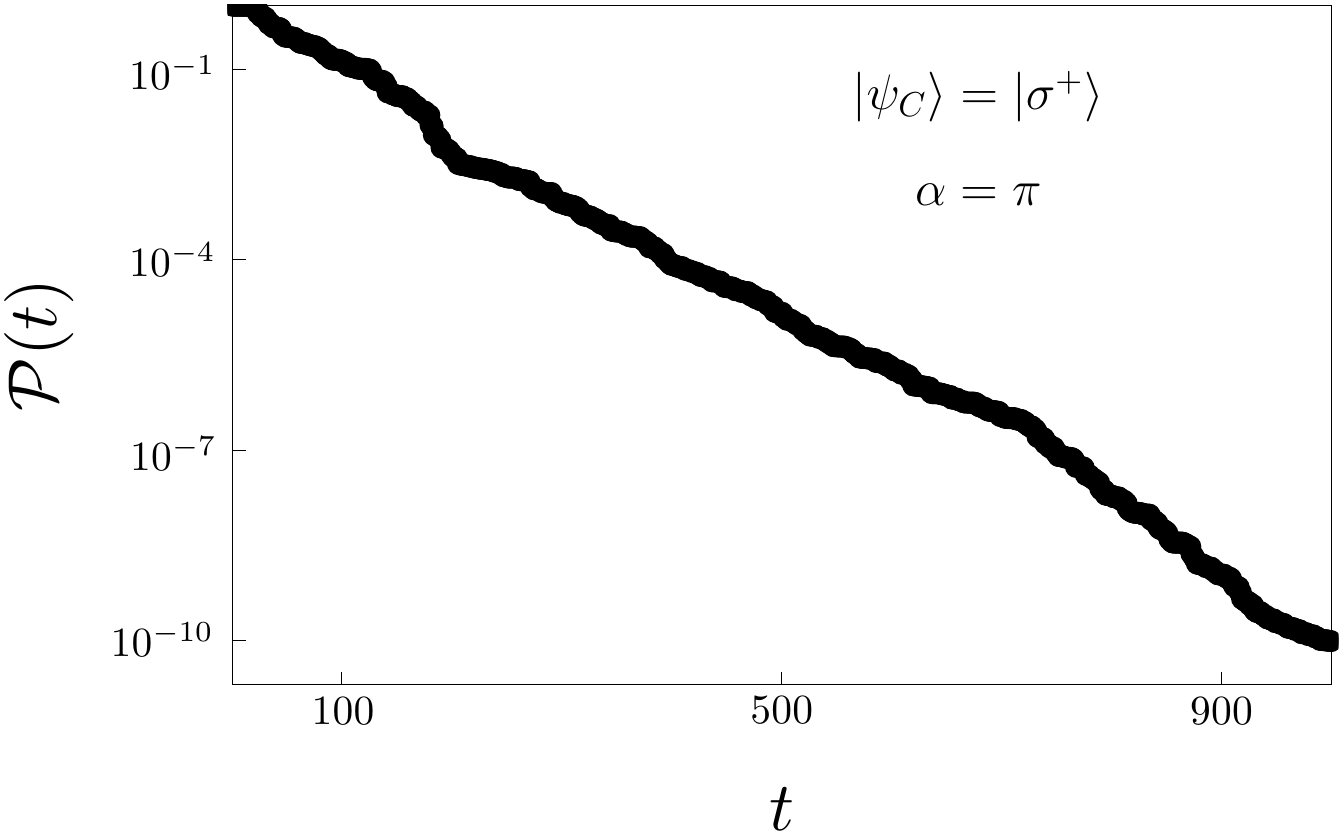}
\end{center}
\caption{Survival probability for percolated quantum walk. The
initial coin state was chosen as $|\sigma^+\rangle$. The probability
of edge presence is $p=\frac{1}{2}$. On the left we have considered
the phase $\alpha=0$. In this case, percolations do not eliminate the trapping effect and the survival probability does not
drop below the same value as for the non-percolated walk (see the right plot of
Figure~\ref{fig:3state}). On the other hand, for $\alpha=\pi$, which
we display in the right plot, the trapping effect vanishes. The
survival probability decreases exponentially, which we highlight
with the log-scale. The deviations from the straight line stem from the fact that the plot corresponds to a single random realization of the percolated walk.}
\label{fig:perc}
\end{figure}

For illustration we display in Figure~\ref{fig:perc} the survival
probability for one random realization of dynamically percolated
quantum walk. On the left we have chosen $\alpha=0$, for which the
stationary states (\ref{stat:state}) are unaffected by percolations.
We find that the survival probability levels at the same value as
for the ideal walk (see the right plot of Figure~\ref{fig:3state}
for comparison). The plot on the right shows the survival
probability when $\alpha=\pi$. In this case percolations cancel the
trapping effect and the survival probability decays exponentially.

The decay rate $\gamma$ of the survival probability depends on both
parameters of the coin $\rho$ and $\alpha$ and also on the
probability of edge presence $p$. The numerical simulations
indicates that the decay rate can be estimated according to
\begin{equation}
\label{gamma:perc}
\gamma = 1-|\lambda_l|,
\end{equation}
where $\lambda_l$ is the leading eigenvalue of the superoperator
\begin{equation}
\label{sup:op}
\Phi = \sum\limits_{\mathcal K} p_{\mathcal K} \left(\hat\pi\hat U_{\mathcal K}\right)\otimes \left(\hat\pi\hat U_{\mathcal K}^*\right),
\end{equation}
which describes the evolution of the density matrix consisting of
the random unitary channel (\ref{ruch}) and the projection onto the
sink (\ref{rho:sink}). Compared to Eq.~(\ref{2state:decay}) the factor 2 is missing due to the use of the superoperator formalism. In (\ref{sup:op}) star denotes the complex conjugation.

\begin{figure}[h]
\begin{center}
\includegraphics[width=0.6\textwidth]{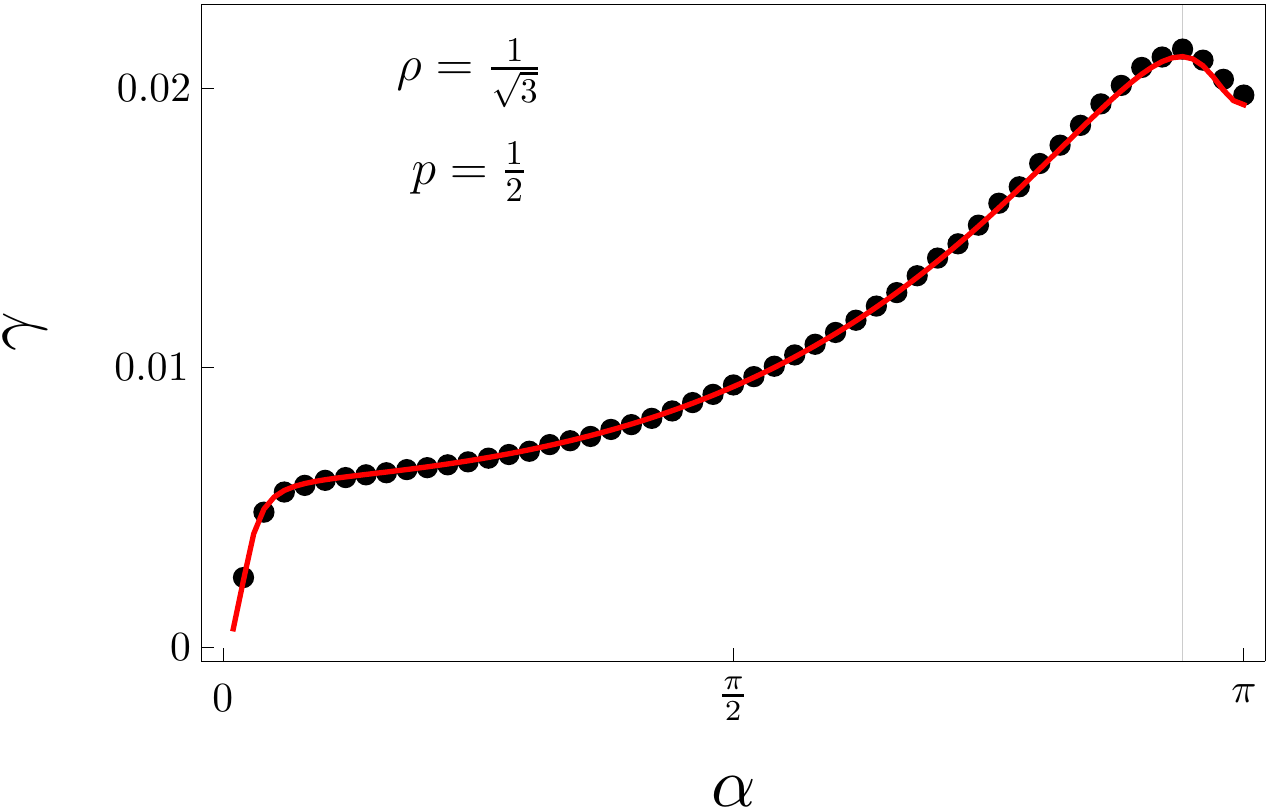}\hfill
\end{center}
\caption{Decay rate as a function of the phase $\alpha$. The other
parameters have been chosen as $\rho=\frac{1}{\sqrt{3}}$ and
$p=\frac{1}{2}$. The maximal decay rate is reached for
$\alpha\approx \frac{47}{50} \pi$.} \label{fig:gamma:phase}
\end{figure}

\begin{figure}[h]
\begin{center}
\includegraphics[width=0.6\textwidth]{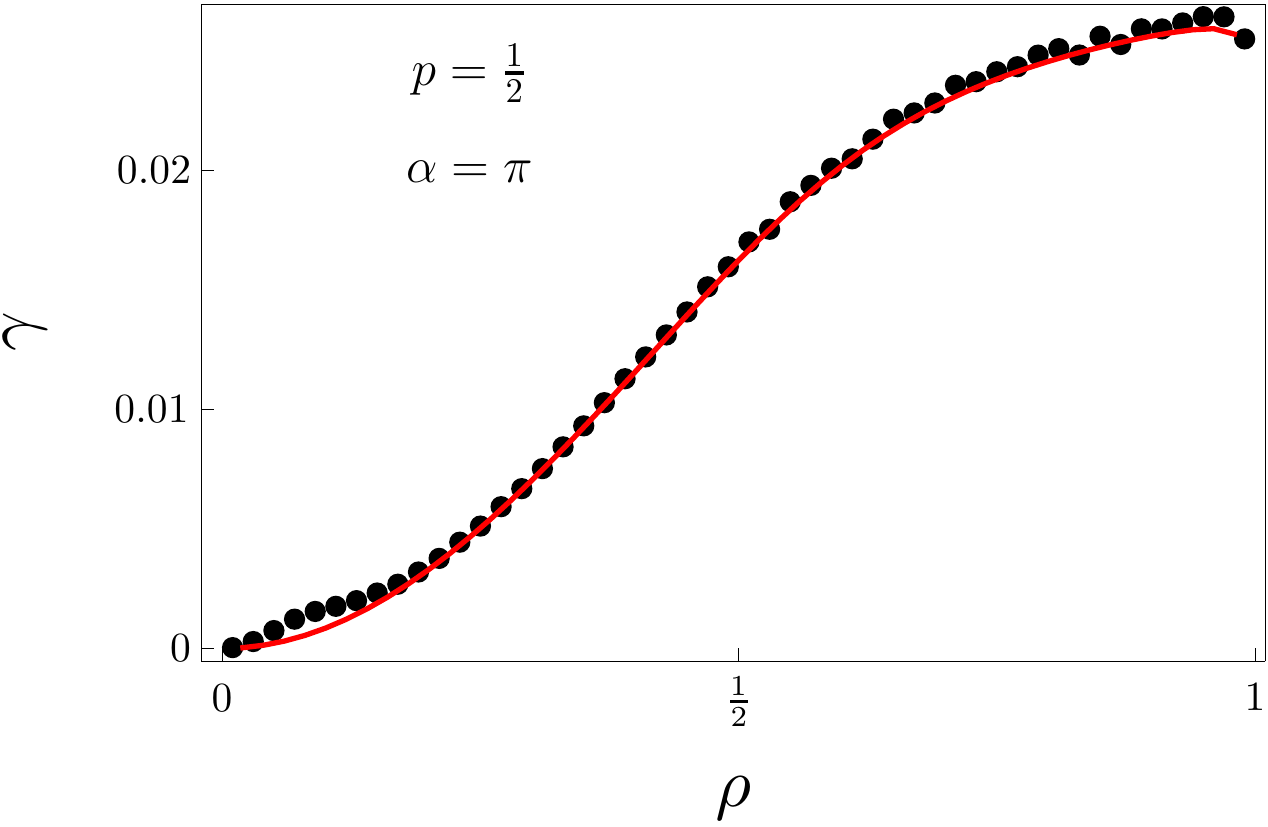}\hfill
\end{center}
\caption{Decay rate as a function of the probability of the coin parameter $\rho$. We have considered $p=\frac{1}{2}$ and $\alpha=\pi$. }
\label{fig:gamma:rho}
\end{figure}

\begin{figure}[h]
\begin{center}
\includegraphics[width=0.6\textwidth]{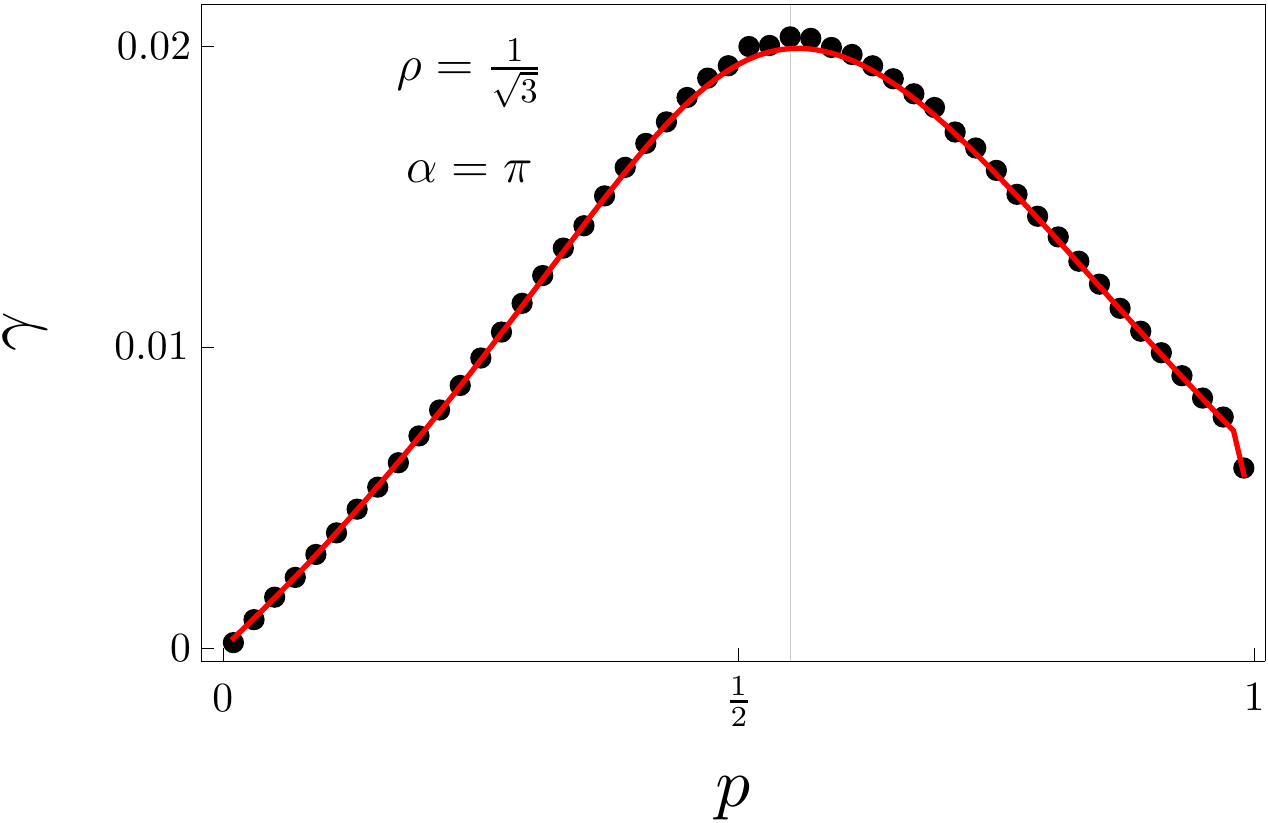}\hfill
\end{center}
\caption{Decay rate as a function of the probability of edge
presence probability $p$ for fixed $\rho=\frac{1}{\sqrt{3}}$ and
$\alpha=\pi$. The maximal decay rate is obtained for $p\approx
0.55$. } \label{fig:gamma:prob}
\end{figure}

For illustration we display in Figures~\ref{fig:gamma:phase},
\ref{fig:gamma:rho} and \ref{fig:gamma:prob} the decay rate as a
function of the phase $\alpha$, coin parameter $\rho$ and the edge
presence probability $p$, respectively. The size of the ring is
given by $N=5$. The red curves are given by the formula
(\ref{gamma:perc}) while the black dots are obtained from numerical
simulation where we fit the exponential decay (\ref{2state:asymp})
to the survival probability averaged over 1000 random realizations
of percolated quantum walk.

In Figure~\ref{fig:gamma:phase} we plot the decay rate as a function
of the phase $\alpha$ while fixing the coin parameter
$\rho=\frac{1}{\sqrt{3}}$ and the edge presence probability
$p=\frac{1}{2}$. For small values of $\alpha$ the decay rate tends
to zero, as can be expected. Notice that maximal decay rate is not
obtained for $\alpha=\pi$, but rather for $\alpha\approx \frac{47}{50}
\pi$. We have not found a simple explanation
for this effect. The numerical simulations indicate that the position of the peak drifts further away from $\frac{\pi}{2}$, however, very mildly. For $\alpha\in(\pi,2\pi)$ the plot would be the mirror image of the presented one.

The decay rate in dependence on the coin parameter $\rho$ is
displayed in Figure~\ref{fig:gamma:rho}. The remaining parameters
were chosen as $p=\frac{1}{2}$ and $\alpha=\pi$.

Figure~\ref{fig:gamma:prob} shows the decay rate as a function of
the edge presence probability $p$ for fixed
$\rho=\frac{1}{\sqrt{3}}$ and $\alpha=\pi$. Notice the asymmetry of
the curve. The maximal decay rate is reached for $p\approx 0.55$. The numerical simulations indicate that with increasing $N$ the position of the maximum tends to $p=0.5$

In summary, percolations eliminate the trapping effect provided that the coin parameter $\alpha$ is non-zero. This leads to exponential decay of the survival probability with the decay rate determined by the leading eigenvalue of the superoperator (\ref{sup:op}). However, for $\alpha=0$ the trapping effect is robust to percolations and the transport efficiency is not improved.

\section{Conclusions}
\label{sec:5}

We have analyzed the absorption problem for discrete-time quantum
walks on a ring. Using both numerical as well analytic methods we
determined several properties of this model of transport. For a two-state
quantum walk the transport of excitation to the sink is efficient
and the survival probability decays exponentially independent of the
initial coin state. The decay rate is determined by the coin
operator and the size of the ring. In this respect we
completed the analysis presented previously.

Next, we have considered a two-parameter set of lazy quantum
walk which exhibits the trapping effect. Compared to the two state quantum walk the lazy walk shows a much richer dynamics. Indeed, the survival probability
has a non-vanishing lower bound and the excitation transport is
inefficient, except for a particular initial coin state. We have
determined the dependency of the transport efficiency on the initial
coin state, the coin operator and the size of the ring.

Finally, we have shown that the trapping effect can be eliminated by
dynamical percolations of the ring provided that the phase parameter
$\alpha$ of the coin operator is non-zero. In such a case, the
survival probability decays exponentially independent of the initial
condition. The decay rate is determined by the parameters of the
coin and percolations. However, for $\alpha=0$ the stationary states
are resilient to percolations and the trapping effect is preserved. We note that in the framework of continuous-time quantum walks similar effects have been found in \cite{Muelken:1,Muelken:2,Muelken:3,Darazs:fractal}.

The trapping effect is present also in quantum walks on more
complicated graphs driven by higher dimensional coins. It would be interesting to find conditions under which it is robust under percolations,  or, on the contrary, what type of percolation is sufficient to eliminate the trapping effect and allow for efficient transfer. In this way studies of percolated quantum walk could contribute to our understanding of transport along complicated molecular structures and ways how to control it.

Finally, let us briefly comment on the possible physical implementations of the lazy walk model. Since the model requires three internal states, usual optical implementations based on polarization are not applicable, at least not in a straightforward way. However, one may employ optical angular momentum \cite{craig:oam} or interferometric multiports \cite{hillery:sqw}. Additional candidates might be realizations of quantum walk in phase space \cite{xue:phase:space1,xue:phase:space2} or using trapped three-level atoms \cite{3level:atom,3level:rydberg}.

\ack

We acknowledge the financial support from RVO 68407700. M\v S is grateful for the financial support from GA\v CR 14-02901P. IJ is grateful for the financial support from GA\v CR 13-33906S.

\appendix

\section{Survival probability for a two-state walk}
\label{app:2state}

In this appendix we show that the survival probability for a two-state quantum walk model which we have discussed in Section~\ref{sec:2} is independent of the initial coin state. The key ingredients of the proof are the results of \cite{konno:absorp} and the fact that we consider a highly symmetric situation. Namely, the excitation enters the ring exactly opposite of the sink.

In \cite{konno:absorp} the authors have studied the absorption problem for a two-state quantum walk on a finite line with vertices $\{0,\ldots,n\}$ with sinks on both ends 0 and $n$. In particular, they have focused on the probability $P^{(n)}_k(t,\psi_C)$ that the excitation starting the walk at the vertex $k$ with the initial state $\psi_C=\left(\psi_L,\psi_R\right)^T$ is absorbed at the vertex $0$ after $t$ steps of the walk. By $\psi_{L,R}$ we have denoted the amplitudes of the initial coin state in the standard basis, i.e.
$$
|\psi_C\rangle = \psi_L|L\rangle + \psi_R|R\rangle.
$$
It is straightforward to see that the survival probability ${\cal P}(t)$ for the two-state walk on a ring of size $2N$ can be written equivalently as
\begin{equation}
\label{app:surv:p}
{\cal P}(t) = 1- \left(P^{(2N)}_N(t,\psi_C) + \tilde{P}^{(2N)}_N(t,\psi_C)\right),
\end{equation}
where we have denoted by $\tilde{P}^{(2N)}_N(t,\psi_C)$ the probability of absorption at the vertex $2N$. We now prove that the sum $P^{(2N)}_N(t,\psi_C) + \tilde{P}^{(2N)}_N(t,\psi_C)$ is independent of the initial coin state $\psi_C$. It was shown in \cite{konno:absorp} that the probability of absorption at 0 can be expressed in the form
\begin{equation}
\label{app:absorp:0}
P^{(2N)}_N(t,\psi_C) = C_1(t)|\psi_L|^2 + C_2(t)|\psi_R|^2 + 2{\rm Re}\left(C_3(t)\psi_L^*\psi_R\right).
\end{equation}
The coefficients $C_i(t)$ are determined by the coin operator. For the choice of the coin (\ref{coin:2state}) they read
\begin{eqnarray}
\nonumber C_1(t) & = & \left|\rho p_N^{(2N)}(t) + \sqrt{1-\rho^2}r_N^{(2N)}(t)\right|^2, \\
\nonumber C_2(t) & = & \left|\sqrt{1-\rho^2} p_N^{(2N)}(t) - \rho r_N^{(2N)}(t)\right|^2, \\
\nonumber C_3(t) & = & \left(\rho p_N^{(2N)}(t) + \sqrt{1-\rho^2}r_N^{(2N)}(t)\right)^* \left(\sqrt{1-\rho^2} p_N^{(2N)}(t) - \rho r_N^{(2N)}(t)\right).
\end{eqnarray}
The quantities $p_N^{(2N)}(t)$ and $r_N^{(2N)}(t)$ were analyzed in \cite{konno:absorp}. They also depend on the coin operator and for the choice of the coin (\ref{coin:2state}) they are real valued. Hence, we can omit the complex conjugation in the formula for $C_3(t)$ since all terms involved are real.

Let us now turn to the probability of absorption at the vertex $2N$. This was not considered in \cite{konno:absorp}, however, it is straightforward to map it to the probability of absorption at the vertex 0. Indeed, by interchanging the coin states $|L\rangle$ and $|R\rangle$ we can express $\tilde{P}^{(2N)}_N(t,\psi_C)$ as the probability of absorption at 0 in a quantum walk with the coin operator
$$
\tilde{C}^{(2)} = \left(
                    \begin{array}{cc}
                      -\rho & \sqrt{1-\rho^2} \\
                      \sqrt{1-\rho^2} & \rho \\
                    \end{array}
                  \right),
$$
starting with the initial coin state
$$
|\tilde{\psi}_C\rangle = \psi_R|L\rangle + \psi_L|R\rangle.
$$
Here we also use the symmetry of the problem, i.e. the fact that the distance from the starting point of the walk to sinks at 0 and $2N$ is the same. Modifying the formula (\ref{app:absorp:0}) accordingly we find that the probability of absorption at the vertex $2N$ reads
$$
\tilde{P}^{(2N)}_N(t,\psi_C) = \tilde{C}_1(t)|\psi_R|^2 + \tilde{C}_2(t)|\psi_L|^2 + 2{\rm Re}\left(\tilde{C}_3(t)\psi_L\psi_R^*\right),
$$
with coefficients $\tilde{C}_i(t)$ given by
\begin{eqnarray}
\nonumber \tilde{C}_1(t) & = & \left|-\rho \tilde{p}_N^{(2N)}(t) + \sqrt{1-\rho^2}\tilde{r}_N^{(2N)}(t)\right|^2, \\
\nonumber \tilde{C}_2(t) & = & \left|\sqrt{1-\rho^2} \tilde{p}_N^{(2N)}(t) + \rho \tilde{r}_N^{(2N)}(t)\right|^2, \\
\nonumber \tilde{C}_3(t) & = & \left(-\rho \tilde{p}_N^{(2N)}(t) + \sqrt{1-\rho^2}\tilde{r}_N^{(2N)}(t)\right) \left(\sqrt{1-\rho^2} \tilde{p}_N^{(2N)}(t) + \rho \tilde{r}_N^{(2N)}(t)\right).
\end{eqnarray}
Following \cite{konno:absorp} we find that the quantities $\tilde{p}_N^{(2N)}(t), \tilde{r}_N^{(2N)}(t)$ are related to $p_N^{(2N)}(t),r_N^{(2N)}(t)$ through the formula
\begin{eqnarray}
\nonumber \tilde{p}_N^{(2N)}(t) & = & (-1)^{N-1}p_N^{(2N)}(t),\\
\nonumber \tilde{r}_N^{(2N)}(t) & = & (-1)^N r_N^{(2N)}(t).
\end{eqnarray}
It is then straightforward to show that
\begin{eqnarray}
\nonumber \tilde{C}_1(t) & = & C_1(t),\\
\nonumber \tilde{C}_2(t) & = & C_2(t),\\
\nonumber \tilde{C}_3(t) & = & - C_3(t).
\end{eqnarray}
Hence, the sum of the probabilities of absorption at vertices 0 and $2N$ reads
\begin{eqnarray}
\label{app:sum:p}
\nonumber P^{(2N)}_N(t,\psi_C) + \tilde{P}^{(2N)}_N(t,\psi_C) & = & C_1(t)+C_2(t) + \\
& & + 2{\rm Re}\left(C_3(t)\left(\psi_L^*\psi_R - \psi_L\psi_R^*\right)\right),
\end{eqnarray}
where we have used the normalization condition of the initial coin state
$$
|\psi_L|^2 + |\psi_R|^2 = 1.
$$
Moreover, since $C_3(t)$ is real and $\psi_L^*\psi_R - \psi_L\psi_R^*$ is purely
imaginary the last term in (\ref{app:sum:p}) vanishes. Hence, we find that (\ref{app:sum:p}) is independent of the initial coin state and through the relation (\ref{app:surv:p}) the same holds for the survival probability ${\cal P}(t)$. This completes our proof.



\section*{References}

\begin{thebibliography}{x}

\bibitem{adz}
Aharonov Y, Davidovich L and Zagury N 1993 {\it Phys. Rev.} A {\bf 48} 1687

\bibitem{meyer}
Meyer D 1996 {\it J. Stat. Phys.} {\bf 85} 551

\bibitem{fg}
Farhi E and Gutmann S 1998 {\it Phys. Rev.} A {\bf 58} 915

\bibitem{skw}
Shenvi N, Kempe J and Whaley K 2003 {\it Phys. Rev.} A {\bf 67} 052307

\bibitem{childs:search:2004}
Childs A M and Goldstone J 2004 {\it Phys. Rev.} A {\bf 70} 022314

\bibitem{reitzner:search}
Reitzner D, Hiller M, Feldman E and Bu\v zek V 2009 {\it Phys. Rev.} A {\bf 79} 012323

\bibitem{vasek:search}
Poto\v cek V, Gabris A, Kiss T and Jex I 2009 {\it Phys. Rev.} A {\bf 79} 012325

\bibitem{childs:search:2014}
Childs A M and Ge T 2014 {\it Phys. Rev.} A {\bf 89} 052337

\bibitem{kendon:qw:pst}
Kendon V M and Tamon C 2011 {\it J. Comput. Theor. Nanosc.} {\bf 8} 422

\bibitem{wojcik:qw:pst}
Kurzynski P and Wojcik A 2011 {\it Phys. Rev.} A {\bf 83} 062315

\bibitem{barr:qw:pst}
Barr K E, Proctor T J, Allen D and Kendon V M 2014 {\it Quantum Inf. Comput.} {\bf 14} 417

\bibitem{zhan:qw:pst}
Zhan X, Qin H, Bian Z H, Li J and Xue P 2014 {\it Phys. Rev.} A {\bf 90} 012331

\bibitem{gedik:qw:pst}
Yalcinkaya I and Gedik Z 2015 {\it J. Phys.} A {\bf 48} 225302

\bibitem{gamble}
Gamble J K, Friesen M, Zhou D, Joynt R and Coppersmith S N 2010 {\it Phys. Rev.} A {\bf 81} 052313

\bibitem{berry}
Berry S D and Wang J B 2011 {\it Phys. Rev.} A {\bf 83} 042317

\bibitem{rudiger}
Rudinger K, Gamble J K, Wellons M, Bach E, Friesen M, Joynt R and Coppersmith S N 2012 {\it Phys. Rev.} A {\bf 86} 022334

\bibitem{reitzner}
Hillery M, Reitzner D and Bu\v zek V 2010 {\it Phys. Rev.} A {\bf 81} 062324

\bibitem{hillery}
Hillery M, Zheng H J, Feldman E, Reitzner D and Bu\v zek V 2012 {\it Phys. Rev.} A {\bf 85} 062325

\bibitem{cottrell}
Cottrell S and Hillery M 2014 {\it Phys. Rev. Lett.} {\bf 112} 030501

\bibitem{childs}
Childs A M 2009 {\it Phys. Rev. Lett.} {\bf 102} 180501

\bibitem{Lovett}
Lovett N B, Cooper S, Everitt M, Trevers M and Kendon V 2010 {\it Phys. Rev.} A
{\bf 81} 042330

\bibitem{karski}
Karski M,  F{\"o}rster L, Choi J, Steffen A, Alt W, Meschede D and Widera A 2009 {\it Science} \textbf{325} 174

\bibitem{schmitz}
Schmitz H, Matjeschk R, Schneider C, Glueckert J, Enderlein M, Huber T and Schaetz T 2009 {\it Phys. Rev. Lett.} \textbf{103} 090504

\bibitem{zahringer}
Z{\"a}hringer F, Kirchmair G, Gerritsma R, Solano E, Blatt R and Roos C F 2010 {\it Phys. Rev. Lett.} \textbf{104} 100503

\bibitem{and:1d}
Schreiber A, Cassemiro K N, Poto{\v c}ek V, G\'abris A, Mosley P J, Andersson E, Jex I and Silberhorn C 2010 {\it Phys. Rev. Lett.} \textbf{104} 050502

\bibitem{broome}
Broome M A, Fedrizzi A, Lanyon B P, Kassal I, Aspuru-Guzik A and White A G 2010 {\it Phys. Rev. Lett.} \textbf{104} 153602

\bibitem{peruzzo:corelated:photons}
Peruzzo A, Lobino M, Matthews J C F, Matsuda N, Politi A, Poulios K, Zhou X, Lahini Y, Ismail N, Worhoff K, Bromberg Y, Silberberg Y, Thompson M G and O'Brien J L 2010 {\it Science} {\bf 3329} 1500

\bibitem{two:photon:waveguide}
Owens J O, Broome M A, Biggerstaff D N, Goggin M E, Fedrizzi A, Linjordet T, Ams M, Marshall g D, Twamley J, Withford M J and White A G 2011 {\it New J. Phys.} {\bf 13} 075003

\bibitem{sansoni}
Sansoni L, Sciarrino F, Vallone G, Mataloni P, Crespi A, Ramponi R and Osellame R 2012 {\it Phys. Rev. Lett.} \textbf{108} 010502

\bibitem{and:2dwalk:science}
Schreiber A, G\'abris A, Rohde P P, Laiho K, {\v S}tefa{\v n}\'ak M, Poto{\v c}ek V, Hamilton C, Jex I and Silberhorn C 2012 {\it Science} \textbf{336} 55

\bibitem{delayed:choice}
Jeong Y C, Di Franco C, Lim H T, Kim M S and Kim Y H 2013 {\it Nature Comm.} {\bf 4} 2471

\bibitem{elster:perc}
Elster F, Barkhofen S, Nitsche T, Novotn\'y J, G\'abris A, Jex I and Silberhorn C 2015 {\it Sci. Rep.} {\bf 5} 13495

\bibitem{ambainis:absorp}
Ambainis A, Bach E, Nayak A, Vishwanath A and Watrous J 2001 in {\it Proceedings of the Thirty-Third Annual ACM Symposium on Theory of Computing} 37

\bibitem{konno:absorp}
Konno N, Namiki T, Soshi T and Sudbury A 2003 {\it J. Phys.} A {\bf 36} 241

\bibitem{bach:absorp}
Bach E, Coppersmith S, Goldschen M P, Joynt R and Watrous J 2004 {\it J. Comput. System Sci.} {\bf 69} 562

\bibitem{yamasaki:absorp}
Yamasaki T, Kobayashi H and Imai H 2003 {\it Phys. Rev.} A {\bf 68} 012302

\bibitem{kwek:absorp}
Kwek L C and Setiawan 2011 {\it Phys. Rev.} A {\bf 84} 032319

\bibitem{chandra:absorp}
Chandrashekar C M and Busch T 2014 {\it Quantum Inf. Proc.} {\bf 13} 1313

\bibitem{janos:absorp}
Asboth J K and Edge J M 2015 {\it Phys. Rev.} A {\bf 91} 022324

\bibitem{inui:grover1}
Inui N, Konno N and Segawa E 2005 {\it Phys. Rev.} E \textbf{72} 056112

\bibitem{inui:grover2}
Inui N and Konno N 2005 {\it Physica} A \textbf{353} 133

\bibitem{stef:cont:def}
\v Stefa\v n\'ak M, Bezd\v ekov\'a I and Jex I 2012 {\it Eur. Phys. J.} D \textbf{66} 142

\bibitem{stef:3state}
\v Stefa\v n\'ak M, Bezd\v ekov\'a I, Jex I and Barnett S M 2014  {\it Quantum Inf. Comput.} {\bf 14} 1213

\bibitem{grimmett:perc}
Grimmett G 1999 {\it Percolation, Die Grundlehren der mathematischen Wissenschaften in Einzeldarstellungen} (Springer, New York)

\bibitem{kendon:perc}
Leung G, Knott P, Bailey J and Kendon V 2010 {\it New J. Phys.} {\bf 12} 123018

\bibitem{unitary:equivalence}
Goyal S K, Konrad T and Diosi L 2015 {\it Phys. Lett.} A {\bf 379} 100

\bibitem{kempf}
Kempf A and Portugal R 2009 {\it Phys. Rev.} A {\bf 79} 052317

\bibitem{stef:eigen}
\v Stefa\v n\'ak M, Bezd\v ekov\'a I and Jex I 2014 {\it Phys. Rev.} A {\bf 90} 012342

\bibitem{balint:perc1}
Koll\'ar B, Kiss T, Novotn\'y J and Jex I 2012 {\it Phys. Rev. Lett.} {\bf 108} 230505

\bibitem{balint:perc2}
Koll\'ar B, Novotn\'y J, Kiss T, and Jex I 2014 {\it Eur. Phys. J. Plus} {\bf 129} 103

\bibitem{Muelken:1}
Muelken O and Blumen A 2011 {\it Phys. Rep.} {\bf 502} 37

\bibitem{Muelken:2}
Schijven P, Kohlberger J, Blumen A and Muelken O 2012 {\it J. Phys.} A {\bf 45} 215003

\bibitem{Muelken:3}
Anishchenko A, Blumen A and Muelken O 2013 {\it Phys. Rev.} E {\bf 88} 062126

\bibitem{Darazs:fractal}
Dar\'azs Z, Anishchenko A, Kiss T, Blumen A and Muelken O 2014 {\it Phys. Rev.} E {\bf 90} 032113

\bibitem{craig:oam}
Hamilton C S, G\'abris A, Jex I and Barnett S M 2011 {\it New J. Phys.} {\bf 13} 013015

\bibitem{hillery:sqw}
Hillery M, Bergou J A and Feldman E 2003 {\it Phys. Rev.} A {\bf 68} 032314

\bibitem{xue:phase:space1}
Xue P, Sanders B C, Blais A and Lalumiere K 2008 {\it Phys. Rev.} A {\bf 78} 042334

\bibitem{xue:phase:space2}
Xue P and Sanders B C 2008 {\it New J. Phys.} {\bf 10} 053025

\bibitem{3level:atom}
Eckert K, Mompart J, Corbalan R, Lewenstein M and Birkl G 2006 {\it Optics Commun.} {\bf 264} 264

\bibitem{3level:rydberg}
Sevincli S et al. 2011 {\it J. Phys.} B {\bf 44} 184018

\end{thebibliography}
\end{document}